\documentclass[aps,prl,twocolumn,superscriptaddress]{revtex4-1}
\usepackage{epsfig}
\usepackage[dvipsnames,usenames]{color}
\usepackage{amsmath}
\usepackage{xfrac}
\emergencystretch=\maxdimen

\usepackage{mathrsfs}
\usepackage{mathtools}
\usepackage{amsfonts}
\usepackage{amsthm}
\usepackage{amssymb}

\DeclareMathOperator{\tr}{Tr}

\newcommand{\be}{\begin{equation}}
\newcommand{\ee}{\end{equation}}

\renewcommand{\[}{\left[}

\usepackage{bm}% bold math

\begin{document}

\title{Measuring von Neumann entanglement entropies without wave functions}

\author{T. Mendes-Santos}
\email{tmendes@ictp.it}
\affiliation{The Abdus Salam International Centre for Theoretical Physics, strada Costiera 11, 34151 Trieste, Italy}
\author{G. Giudici}
\email{ggiudici@ictp.it}
\affiliation{The Abdus Salam International Centre for Theoretical Physics, strada Costiera 11, 34151 Trieste, Italy}
\affiliation{SISSA, via Bonomea 56, 34151 Trieste, Italy}
\affiliation{INFN, sezione di Trieste, 34136 Trieste, Italy}
\author{R. Fazio}
\affiliation{The Abdus Salam International Centre for Theoretical Physics, strada Costiera 11, 34151 Trieste, Italy}
\affiliation{NEST, Scuola Normale Superiore \& Istituto Nanoscienze-CNR, I-56126 Pisa, Italy}
\author{M. Dalmonte}
\affiliation{The Abdus Salam International Centre for Theoretical Physics, strada Costiera 11, 34151 Trieste, Italy}
\affiliation{SISSA, via Bonomea 56, 34151 Trieste, Italy}

\begin{abstract}
We present a method to measure the von Neumann entanglement entropy of ground states of quantum many-body systems which does not require access to the system wave function. The technique is based on a direct thermodynamic study of entanglement Hamiltonians, whose functional form is available from field theoretical insights. The method is applicable to classical simulations such as quantum Monte Carlo methods, and to experiments that allow for thermodynamic measurements such as the density of states, accessible via quantum quenches. We benchmark our technique on critical quantum spin chains, and apply it to several two-dimensional quantum magnets, where we are able to unambiguously determine the onset of area law in the entanglement entropy, the number of Goldstone bosons, and to check a recent conjecture on geometric entanglement contribution at critical points described by strongly coupled field theories.
\end{abstract}

\maketitle

\paragraph{Introduction. - } Over the last twenty years, entanglement has emerged as a paramount tool to characterize quantum wave-functions~\cite{Amico2008,fradkinbook,Eisert2010,Laflorencie2016}. 
A striking example is ground states $|\Psi\rangle$ of many-body systems where, given a spatial bipartition dividing the system into regions $A$ and $B$, the entanglement between $A$ and $B$ is
measured by the von Neumann entropy (VNE):
\begin{equation}
S_A=-\text{Tr}_A\rho_A\ln\rho_A, \quad \rho_A=\text{Tr}_B |\Psi\rangle\langle\Psi |.
\end{equation} 
The VNE remarkably provides a systematic way to connect wave function properties to all operational definitions of entanglement, and is of pivotal importance to both quantum information purposes and as a diagnostic tool in quantum many-body theory. Examples of its relevance include the existence of area laws bounding entanglement in ground state of local Hamiltonians~\cite{Eisert2010}, the sharp characterization of conformal field theories (CFTs) in one-dimension 
(1D)~\cite{Holzhey_1994,Calabrese2004,Calabrese_2009}, topological order~\cite{Kitaev_2006,2006PhRvL..96k0405L} 
and spontaneous symmetry breaking~\cite{Metlitski2011}, and its importance in understanding the complexity of classical simulations~\cite{Schollwock2005}.
However, differently from remarkable theoretical~\cite{Moura_Alves_2004,Daley_2012,Alba:2017aa,Elben_2018,Vermersch_2018} and experimental~\cite{Islam:2015aa,Brydges:2018aa} studies aimed at measuring Renyi entropies, the 
VNE has so far eluded a direct experimental verification, as it requires tomographic access to the system wave function - which becomes quickly impractical beyond few spins. 
The same limitations affect numerical methods such as quantum Monte Carlo (QMC)~\cite{Kaul:2013aa}, which typically cannot sample wave functions, and, thus, cannot compute the VNE.

\begin{figure}[]
{\centering\resizebox*{8.5cm}{!}{\includegraphics*{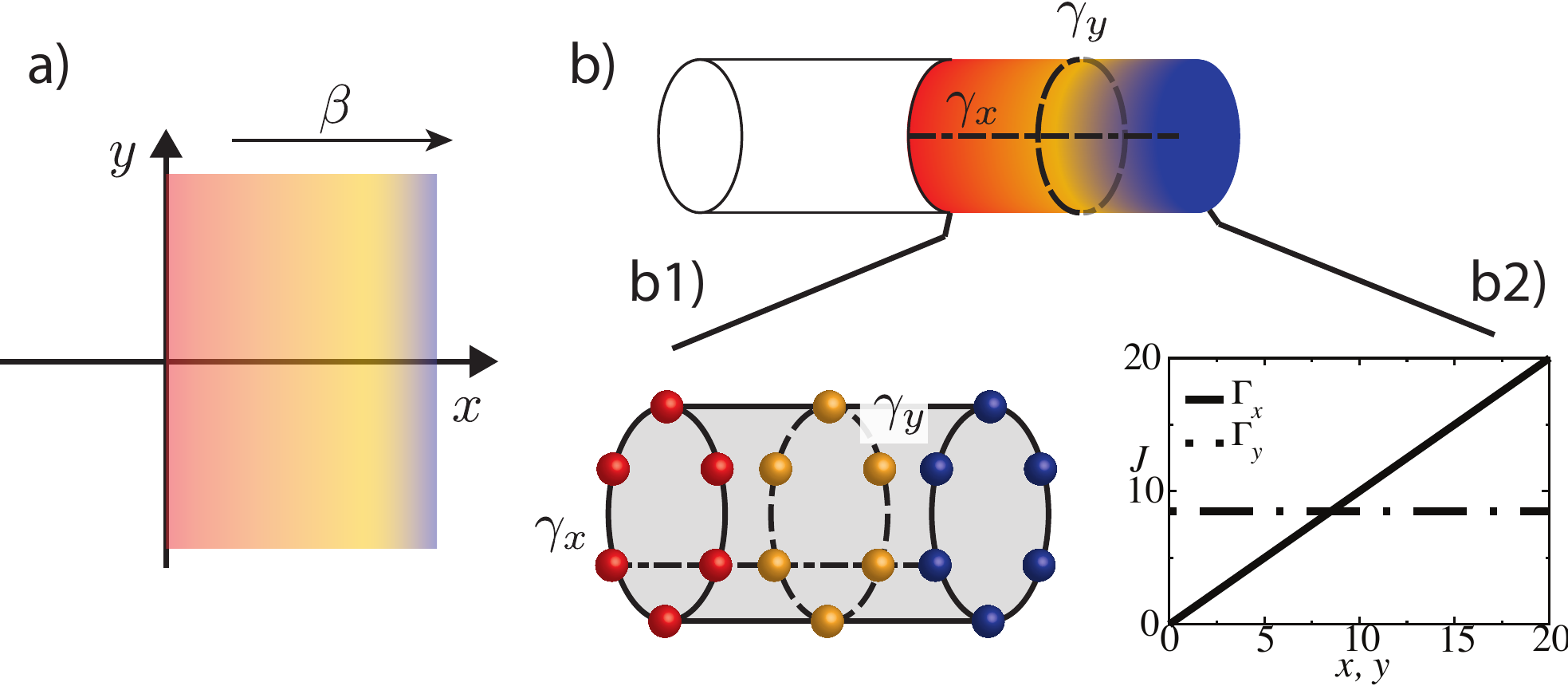}}}
\caption{{\it Entanglement Hamiltonians from field theory to lattice models.} Panel a): schematics of the Bisognano-Wichmann theorem for the 2D case~\cite{bisognano1975duality,bisognano1976duality}. 
A plane is divided into two half-planes at $x=0$. The reduced density matrix of obtained from the vacuum of the field theory upon tracing the $x<0$ region can be interpreted as a thermal equilibrium state with 
inverse temperature $\beta$ increasing as a function of the distance from the boundary. Hot region (red) are typically more entangled then the cold (blue) ones. Panel b): adaption to cylinder geometries. 
In analogy with the infinite plane case, the inverse entanglement temperature is constant at fixed $x$ (path $\gamma_y$), while it increases at fixed $y$ (path $\gamma_x$). 
This picture can immediately be adapted to lattices [panel b1)]~\cite{Dalmonte:2017aa,Giuliano2018}: as depicted in b2), the couplings of the corresponding lattice entanglement
Hamiltonian are constant along $y$ ($\Gamma_y$), while they increase along $x$ ($\Gamma_x$).} 
\label{fig_cart} % Fig 1
\end{figure}

In this work, we describe an approach to measure the von Neumann entanglement entropy of ground states {\it without} relying on probing wave functions, that can be efficiently implemented in both experiments and QMC simulations. 
The backbone of the technique is the formulation of the entanglement measurement problem in terms of the thermodynamic properties of the entanglement (modular)
Hamiltonian (EH) $\tilde{H}_A=-\ln\rho_A$~\cite{bisognano1975duality,bisognano1976duality,Witten:2018aa}, whose structure is schematically depicted in Fig.~\ref{fig_cart}. 
As we show below, our method allows to perform accurate entanglement-based measurements of universal quantitites, such as the number of Nambu-Goldstone modes~\cite{Metlitski2011} 
and central charges~\cite{Holzhey_1994,Calabrese2004}, at the percent level, even for modest system sizes. Most remarkably, it allows the calculation of the entanglement of many-body systems in a scalable 
manner (and well beyond what can be done with alternative numerical methods), thanks to its thermodynamic analogy. 
This is a key point when interested in universal quantities, as those are captured by subleading corrections to the entropy in dimensions $D>1$. 
In terms of experiments, our work is complementary to other techniques that have been proposed to measure either Renyi 
entropies~\cite{Moura_Alves_2004,Daley_2012,Islam:2015aa,Brydges:2018aa,Elben_2018,Vermersch_2018,Abanin:2012aa} and entanglement spectra~\cite{Pichler:2016aa,Dalmonte:2017aa}, 
both in terms of observables and in terms of resources, as we detail below.

After benchmarking our method on 1D examples, we carry out QMC simulations on a series of two-dimensional lattice models and calculate their entanglement entropy for partitions of up to thousand sites - inaccessible to any other controlled numerical method. For the 2D Heisenberg and XY models, we provide direct evidence that {\it (i)} the VNE is constrained by the area law (confirming lower bounds based on Renyi entropies), and {\it (ii)} the number of Goldstone modes can be determined with percent accuracy solely from entanglement properties. For the bilayer Heisenberg model, we study the geometric contribution to the entanglement entropy at its strongly coupled critical point, and verify a recent conjecture on $O(N)$ models~\cite{Sachdev2009}.

\paragraph{Thermodynamics of entanglement Hamiltonians. - } The relation between entanglement and thermodynamic quantities has been widely exploited in the quantum mechanics and field theory literature: an epitome in this context is the Unruh effect~\cite{Unruh:1976aa}, that describes how the vacuum appears as an equilibrium finite temperature state from the point of view of an accelerating observer. In the context of axiomatic field theory, this relation is conveniently expressed by the Bisognano-Wichmann (BW) theorem~\cite{bisognano1975duality,bisognano1976duality,Witten:2018aa}. For a Lorentz invariant theory with Hamiltonian density $H(\vec{x})$, $\vec{x} = (x_1, ..., x_{D})$, in $D$ spatial dimensions, the entanglement Hamiltonian of a half-plane bipartition A defined by $x_1>0$ reads:
\begin{align}
\tilde{H}_A = 2 \pi \int_{\vec{x} \in A} d\vec{x}\left[x_1 H(\vec{x}) \right] + c^{\prime},
\label{BWtheorem}
\end{align}
where $c'$ is a constant that ensures $\text{Tr}_A\rho_A=1$. The BW theorem has been extended to different geometrical partitions in the presence of conformal invariance~\cite{hislop1982,Cardy:2016aa,Casini:2011aa}. 

These results can be cast on a discrete space-time lattice~\cite{Dalmonte:2017aa,Giuliano2018} as follows. 
For the sake of simplicity, let us focus on 1D systems with nearest-neighbor interaction, $h_{n,n+1}$,  and  on-site terms, $l_{n}$; 
the 2D case is discussed in the SM. Up to $c'$, the lattice BW-EH ansatz of a subsytem of length $L$ is
\begin{equation}
 \tilde{H}_{EH} = \beta_{EH} \left[ \sum_{n=1}^{L-1} \Gamma(n) h_{n,n+1} + \Gamma(n-1/2) l_{n} \right],
 \label{EHBW}
\end{equation}
The coefficients $\Gamma$ depend on the geometry of the partition\cite{bisognano1975duality,hislop1982,Cardy:2016aa,Giuliano2018}:
(i) for a half-infinite partition under open boundary conditions (OBC), $\Gamma(n) = n$ [see Fig.~\ref{fig_cart} (b2)]; 
(ii) for subsystem embedded in an infinite system, $\Gamma(n) = n \left(L - n \right) / L $ \cite{hislop1982}, which corresponds to periodic boundary condition (infinite PBC);
and for finite systems (iii) with both PBC, $\Gamma(n) = \frac{L}{ \pi} \sin \left( \frac{  \pi x }{ L } \right)$,
and (iv) OBC, $\Gamma (x) = \frac{2L}{\pi} \sin \left(  \frac{ \pi x }{ 2L } \right)$.
It is straightforward to generalize the 
BW-EH ansatz for a N-dimensional lattice model~\cite{Giuliano2018}: in Fig.~\ref{fig_cart}b,
we schematically illustrate it for the cylinder geometries discussed below;
see the SM for more details  \cite{supmat}.

The overall energy scale of Eq.~\eqref{EHBW} is related to the ``speed of light``, $v$, in the corresponding 
low-energy field theory, $\beta_{EH} = \frac{2\pi}{v}$
and plays the role of an effective inverse temperature. The corresponding reduced density matrix on the lattice and the corresponding thermal entropy read: 
\begin{equation}\label{rhoA}
\rho_{BW} = \frac{e^{-\beta_{EH} H_{EH}} }{Z_{EH}}, \quad S_{BW} = -\text{Tr}\rho_{BW}\ln \rho_{BW}
\end{equation}
where the normalization factor is  interpreted as a partition function $Z_{EH}=\text{Tr}e^{-\beta_{EH} H_{EH} }$~\footnote{We note that, while in principle the entropy is nothing but the expectation value of the EH, the normalization factor $Z_{EH}$, which is not universal, makes this approach hardly applicable.}.

The predictive power of the BW theorem on the lattice goes well beyond the low-lying entanglement spectrum, 
for which analytical and numerical evidence is abundant~\cite{Peschel_1999,peschel2009,Itoyama:1987aa,Nienhuis_2009,Kim_2016,Eisler:2017aa,Dalmonte:2017aa,Peschel2018,1806.08060,Toldin:2018aa,Kosior:2018aa,Giuliano2018}: 
in particular, the entanglement entropy of a lattice system $S_A$ can be evaluated via its thermal equivalent $S_{BW}$ so to determine universal quantities at the percent level even from modest system sizes. 
The accuracy progressively increases with system size, as expected for any field theoretical expectation on the lattice. We remark that the validity of Eq.~\eqref{BWtheorem} relies on the underlying field theory being Lorentz invariant: as such, it is applicable to a broad range of phenomena, including quantum critical phases and points with emergent relativistic description, topological phases, lattice gauge theories, just to name few examples; it cannot be applied to other situations, such as disordered systems or ferromagnets, where low-energy Lorentz invariance is absent.

\paragraph{Measuring entanglement entropy in numerical simulations and experiments. - }

Before discussing the concrete validation examples, we illustrate how to measure VNE in both numerical and real experiments. Both strategies rely on the basic fact that, in analogy with thermodynamics, the VNE is uniquely determined once the density of states of the EH, $D_{EH}$, is known. 

From the point of view of classical computations, this can be achieved using any algorithm based on metadynamics.
Below, we illustrate this by applying the quantum version of the Wang-Landau method performed in the stochastic series expansion (SSE) QMC framework \cite{Wessel2003,Wessel2007}. 
This method allows the direct calculation of the free-energy and the entropy of the BW-EH; see the SM for more details. 
One key point we emphasize is that the method straightforward to implement on a working Wang-Landau code
[only requires to implement an inhomogeneous version of the system Hamiltonian, as Eq. \eqref{EHBW}] \cite{codeALPS}.

From an experimental viewpoint, $D_{EH}$ can be conveniently measured following the procedures proposed in Refs.~\cite{Osborne:aa,Schrodi:2017aa} 
in the context of conventional Hamiltonian operators. The main idea here is that the density of states of a given operator $A$ can be obtained by analysing the quench dynamics 
starting from a randomized set of product states, and evolving according to $A$.
We remark that certain universal quantities are then simply accessible via the density of states itself, 
without the need to evaluate the VNE: an example is the central charge in one-dimensional critical systems~\cite{calabrese2008entanglement}.  

A second way of measuring the VNE experimentally is via the specific heat with respect to the EH: in analogy with thermodynamics, this quantity, when integrated over the temperature range $[1/\beta_{EH},\infty]$ returns the von Neumann entropy. These temperatures are in units of the speed of sound, which can be measured independently via spectroscopy. However, this approach is limited to systems where a controlled exchange with the reservoir is possible, e.g., cold atoms in optical lattices~\cite{Catani:2009aa}.

\begin{figure}[]
{\centering\resizebox*{8.4cm}{!}{\includegraphics*{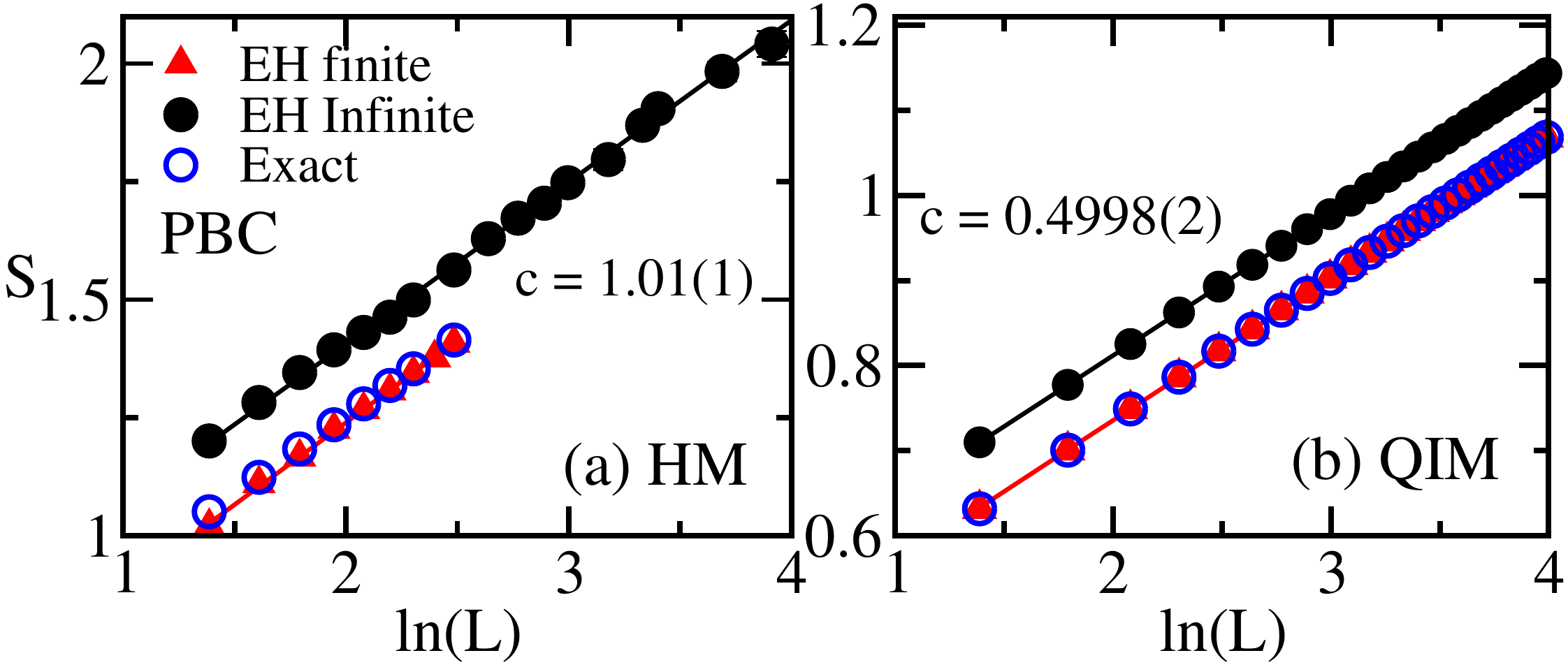}}}
\caption{\textit{BW-EH entropy of one-dimensional critical systems.}
In panels (a) and (b) are shown results for the HM and QIM with PBC, respectively, considering the BW-couplings $\lambda(n)$ (see text).
The central charge obtained from the BW-EH entropy is in agreement with exact results ($c = 1$ and $c = 0.5$). Error bars are smaller than the size of the symbols.} 
\label{fig1} % Fig 1
\end{figure}

\paragraph{One-dimensional critical systems. -} We now benchmark our strategy for one-dimensional critical systems, where the calculation of the VNE is amenable to both exact and tensor network simulations. 
In this case, the VNE of a subsystem of size $L$ diverges logarithmically, 
$S(L) \propto c \ln L$, where $c$ is the central charge of the underlying CFT. 

In Fig.~\ref{fig1}, we plot the BW VNE  of the one-dimensional Heisenberg model (HM)
and the quantum Ising model (QIM) at its quantum critical point, and under PBCs. 
Throughout this work, we employ dimensionless energy units for the sake of convenience.
For the two models, the exact value of the entropy (empty circles) is evaluated using density-matrix-renormalization-group~\cite{White1992} (HM) 
and exact diagonalization methods for a biparition of size $L$ embedded in systems of size $2L$.
The calculations of the BW-EH thermal 
entropy are carried out with QMC with both local and SSE directed-loop updates~\cite{sandvik1991,sandvik2002} for the HM, and exact diagonalization for the QIM. 
In addition to the finite-size EH (red triangles), 
for the sake of comparison, we also compute the entropy obtained utilizing the EH of a finite partition in an infinite system (black circles)~\cite{Giuliano2018}: the two are separated only by a constant shift 
that depends solely on the central charge~\cite{supmat}. 

In both cases, the VNE increases logarithmically as expected: the corresponding central charge considering systems up to $L=80$ $(100)$ is in within 1\% (0.05\%) 
level with the exact results for the HM (QIM) - see Figs.~\ref{fig1} (a) and (b). The difference $\Delta S(L) = S_{BW} - S_{exact}$ goes to zero as $L \to \infty$. 
Remarkably, the BW-EH captures also corrections to the CFT scaling due to marginal operators~\cite{Laflorencie2006,Calabrese2010} under OBCs, as we discuss in the SM.
These results strongly suggest that the EH-BW entropy provides the exact VNE in the limit $L \to \infty$ for one-dimensional critical systems, capturing both universal (leading contribution) 
and non-universal (constant) contributions.

\paragraph{Two-dimensional quantum magnets.}

\begin{figure}[]
{\centering\resizebox*{8.5cm}{!}{\includegraphics*{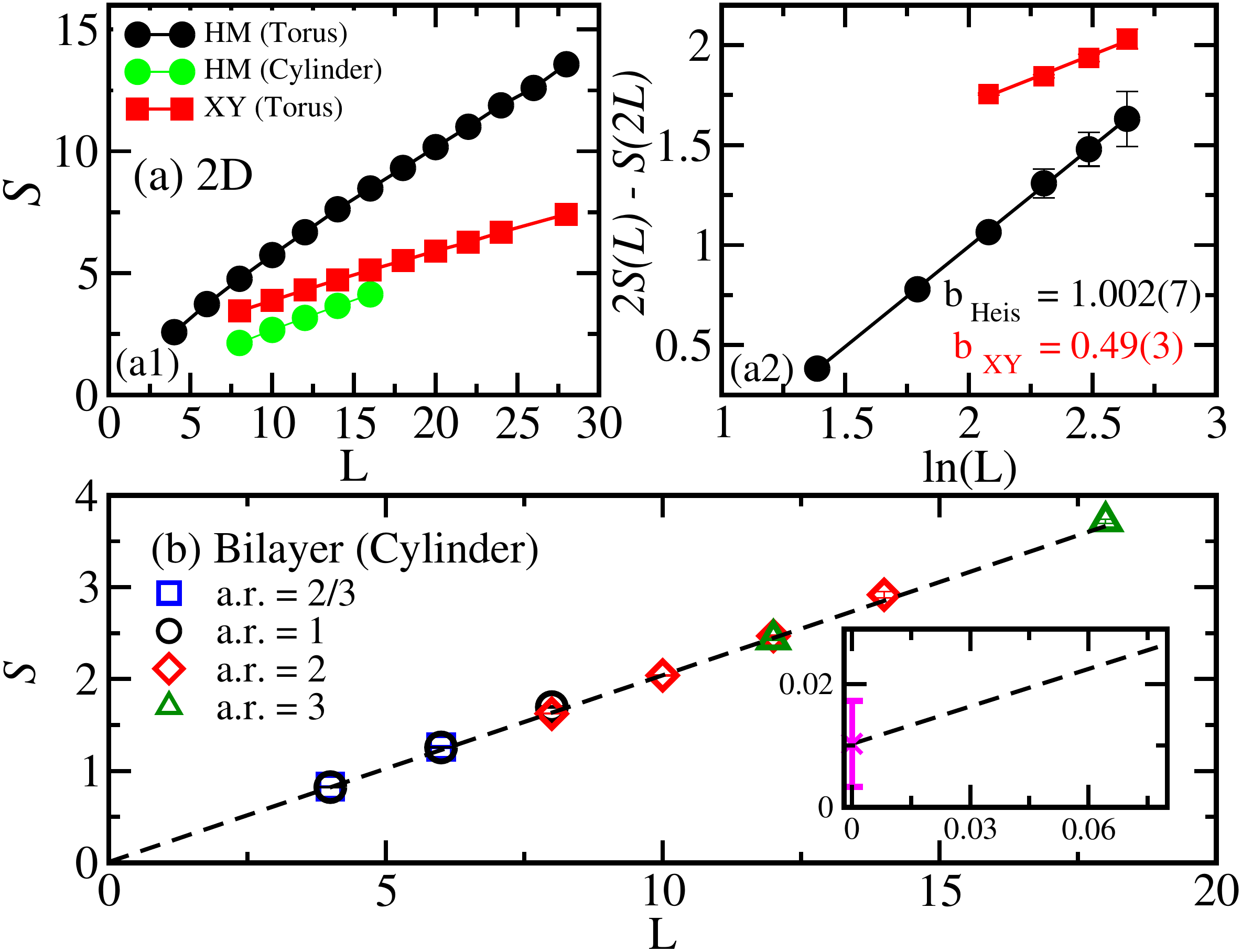}}}
\caption{\textit{BW-EH entropy of two-dimensional systems}. 
Panel (a)  shows results for the HM and XY model. 
The $x$-axis of (a1) represents the linear size of the boundary, $L_y = L$,
and the subsystem aspect ratio for the HM (Torus) is $a.r. = L_y/L_x = 1$, while for the  XY (torus) and the HM (cylinder), $a.r. = 2$.
In panel (a2), we remove the \textit{area law} terms  of $S$, and plot the subleading term of $S$ as function of $\ln L$.
The number of Goldstone modes, $n_b=2b$, extracted with a linear fit, is in agreement with 
expected results.
Panel (b) shows results for the bilayer HM entropy at the QCP, $g_c = 2.522$, and different $a.r.$.
The results are well described by a linear fit, and the $y$-intercept is $\gamma \approx 0$, see the \textit{inset}.} 
\label{fig2} % Fig 1
\end{figure}

The VNE also describes universal properties of two-dimensional systems.
For instance, the VNE of 2D ground states that break a continuous symmetry 
scales as $S(L) = AL + B \ln(L) + D$, where $L$ is the linear size of the boundary.
The $A$ is the non-universal \textit{area law} term~\cite{Eisert2010}, while,
for a smooth boundary, the prefactor of the logarithmic term is a universal quantity related to the number of Nambu-Goldstone modes $n_b$, $B = n_b/2$, 
of the associated spontaneuosly-symmetry-broken (SSB) phase~\cite{Singh2011,Metlitski2011}. 
As examples of SSB, we consider the 2D XY model and the Heisenberg model. 
In both cases, we perform QMC simulations of the EH and extract the corresponding VNE as a function of the subsystem linear size, $L$. 
The entropy is evaluated at $\beta_{EH} = 2 \pi / v$, with $v_{Heis} = 1.658 J $ \cite{sandvik2015} and $v_{XY} = 1.134J$ \cite{sandvik1999}, using the SSE-Wang Landau algorithm.

In Fig.~\ref{fig2}a, we show the scaling of the BW VNE for both cylinder and torus geometries. The scaling is clearly linear. In the case of the HM on a torus, we extracted the coefficient $A$ by fitting these results to $S(L) = AL + B \ln(L) + D$, and obtain $A = 0.372(6) $, which is in agreement with a prediction based on spin-wave approximation~\cite{Laflorencie2011} (discrepancy $< 3\%$).

In Fig. \ref{fig2}b, we extract the subleading logarithmic correction by considering the entropy difference $2S(L)-S(2L)\simeq \frac{n_b\log(L)}{2}$ in toroidal geometries of circumference $2L$. The number of Nambu-Goldstone modes obtained from the prefactor of this term is in perfect agreement with field theoretical expectations~\cite{Metlitski2011,Laflorencie2011,Humeniuk:2012aa,Melko2015}, with accuracy at the percent level or lower. The fact that the VNE returns a value which is considerably closer to the field theoretical prediction when compared to the one extracted from Renyi entropies~\cite{Singh2011,Humeniuk:2012aa} may signal the fact that the latter are more affected by irrelevant operators, as observed in 1D~\cite{Laflorencie2006,Calabrese2010,Dalmonte:2011aa}, or may be due to the smoother continuity properties of the VNE.

\paragraph{Strongly coupled Quantum criticality. -}

As a second example of 2D system, we consider the bilayer Heisenberg model \cite{sandvik1994,sandvik2006}. This model describes a quantum phase transition induced by the inter-coupling $g$
that belongs to the $O(3)$ universality class. We compute the BW-EH entropy at the QCP, $g_c = 2.522$,  considering $\beta_{EH} = 2 \pi / v$, with $v = 1.9001(2) $ \cite{sandvik2015}.
The details of how we cast the BW-EH in the bilayer HM are discussed in Ref.~\cite{supmat}. For this universal class, it has been argued that there is a universal constant correction to the entanglement entropy that depends solely on the aspect ratio~\cite{Sachdev2009}: for PBC, this constant has been conjectured to vanish, in sharp contrast to anti-PBC. Verifying this conjecture requires accurate values of the entropy at large system sizes of several hundred sites.

Our results up to partition of size $L=18$  are depicted in Fig.~\ref{fig2}c. 
Within error bars, our results show that $S(L)$ is independent of the aspect ratio of the subsytem, see Fig. \ref{fig2} (c), 
and have no detectable logarithmic subleading term (the $S(L) = AL + B \ln(L) + D$ fitting, gives $B = -0.05(8)$). 
The $y$-intercept of $S(L)$ is $0.010(7)$, which confirms the conjecture for the universal constant contribution for the $O(N)$ model~\cite{Sachdev2009}.

\begin{figure}[]
{\centering\resizebox*{8.4cm}{!}{\includegraphics*{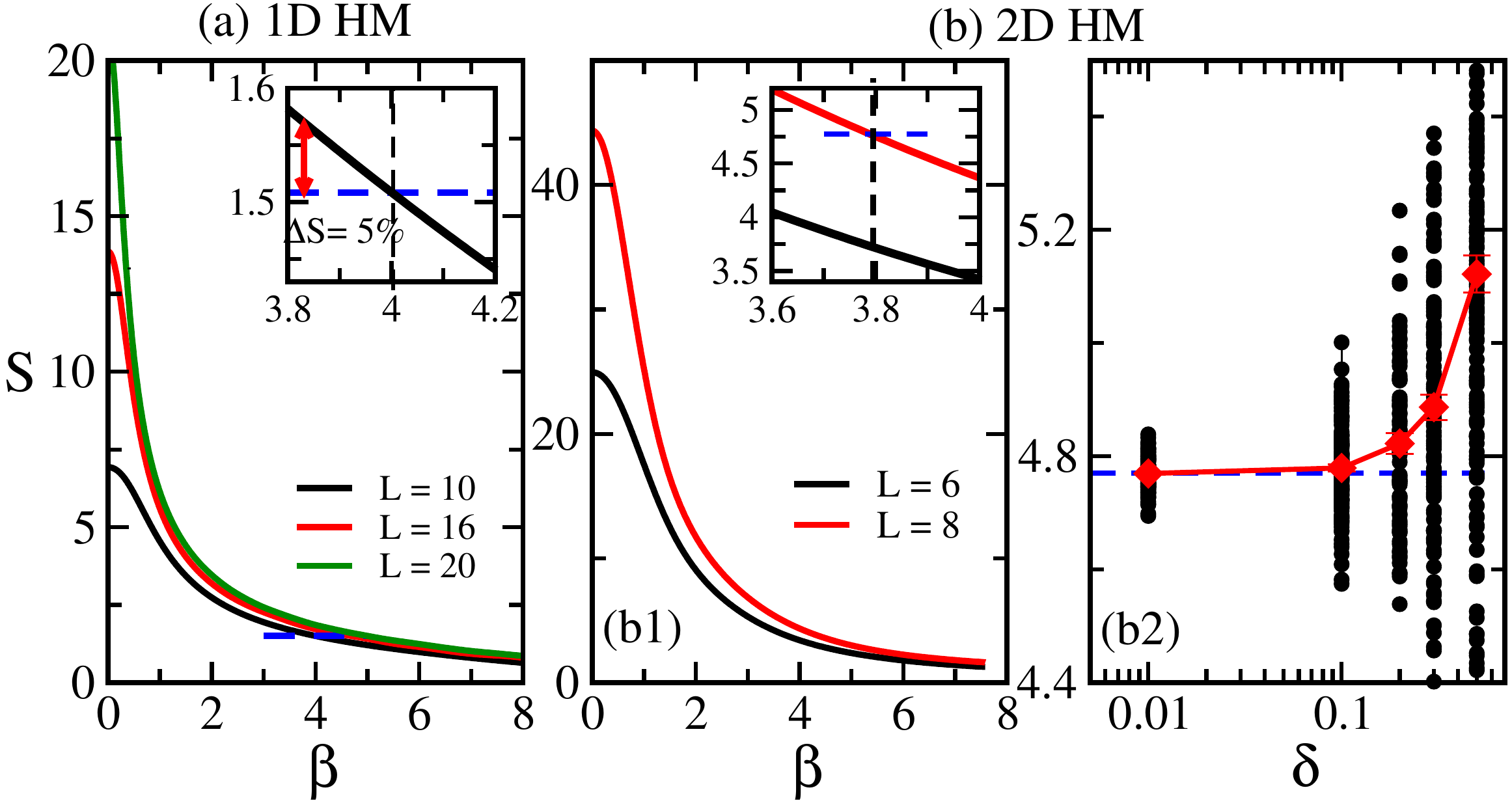}}}
\caption{\textit{Stability to pertubations.}
Panels (a) and (b1) show the $\beta$-dependence of the BW-EH entropy for the 1D and 2D HM, respectively; 
the insets magnify the regions close to the exact value of $\beta_{EH}$ (dashed vertical line). 
In panel (b2)  is shown $S$ as a function of the disorder magnitude $\delta$ for the 2D HM with $L = 8$ (see text).
The circles (black points) are the value of $S$ for a single realization of disorder, while the diamonds (red points) are the averaged BW-EH entropy ($N_r = [100-200]$ realizations of disorder are used).} 
\label{fig4} % Fig 1
\end{figure}

\paragraph{Stability of BW-EH entropy. -} 

Finally, we discuss how stable are the BW-EH entropy to perturbations.
The two main effects we consider are {\it (i)} errors in the value of $\beta_{EH}$, and  {\it (ii)} random perturbations in the EH couplings $\Gamma(n)$. 
The first is relevant for both QMC simulations and experiments, where in general the value
of $\beta_{EH}$  is not known exactly,  while the second accounts for possible imperfect experimental realizations of the EH.

In Fig.\ref{fig4} (a,b1), we show the value of the extracted entropy obtained via Wang-Landau sampling as a function of $\beta$, for both 1D and 2D HM. 
The insets magnify the region in the vicinity of the exact value of $\beta_{EH}$, signaled by a dashed vertical line: in this regime, the entropy is linearly sensitive to $\beta$.
This implies that the accuracy in estimating $S$ is ultimately limited by the accuracy on the sound velocity: this strengthen the applicability of our method to QMC simulations,
where $v$ can be measured very accurately via a variety of techniques~\cite{Kaul:2013aa,sandvik2015}. 

Next, we consider the effect of disordered couplings in the EH, $\Gamma(n) \to \Gamma(n)(1+\delta_n)$, where $\delta_n$ is a random number in the interval $[-\delta,\delta]$~\footnote{We remind that this imperfect EH corresponds to the GS of a clean system, and is not related to the entanglement properties of disordered systems.}. In In Fig.\ref{fig4} (b2), we plot the average value of the entropy as a function of $\delta$ for the 2D HM (the 1D case is discussed in Ref.~\cite{supmat}). The average value of the entropy is very robust to disorder strengths of magnitude 10\%: this remarkable stability is in contrast to what is typically found when studying the effects of disorder in the Hamiltonian couplings, which have a quantitatively larger effect on entropies. A possible element in support of this unexpected resilience is the fact that the VNE is endowed with particularly robust continuity properties with respect to changes in the entanglement spectrum 
(which is instead expected to be directly affected by the random couplings).

\paragraph{Conclusions.} We have presented a method to measure the ground state von Neumann entropy of a broad class of lattice models via direct thermodynamic probe of the correspondent entanglement Hamiltonian. The method is straightforward to implement in quantum Monte Carlo codes, and is of immediate applicability to experiments capable of measuring the density of states. It enables accurate predictions of universal quantities solely based on entanglement, thanks in particular to its immediate scalability in numerical simulations. Future perspectives include the application of the method to other entanglement related quantities, such as the negativity, its extension to lattice gauge theories, and its integration with methods to determine the EH at finite temperature~\cite{Toldin:2018aa,frerot2017}.

\acknowledgements{
We acknowledge useful discussions with C. Gro\ss, N. Laflorencie, A. Laio, M. Lukin, H. Pichler, S. Sachdev, E. Tonni, B. Vermersch, and P. Zoller, and thank P. Calabrese for feedback on the manuscript.
This work is partly supported by the ERC under grant number 758329 (AGEnTh), and has received funding from the European Union's Horizon 2020 research and innovation programme under grant agreement No 817482. 
TMS and MD acknowledge computing resources at Cineca Supercomputing Centre through the
Italian SuperComputing Resource Allocation via the ISCRA grants QMCofEH. 
We used the ALPS to benchmark the results of our code~\cite{Bauer2011}.
TMS and GS equally contributed to this work.}

%%%%%%%%%%%%%%%%%%%%%%%%%%% Bibliography %%%%%%%%%%%%%%%%%%%%%%%%%%%%%%%%

\phantomsection
\addcontentsline{toc}{chapter}{Bibliography}
\bibliography{EntMeasBW}

\begin{thebibliography}{72}%
\makeatletter
\providecommand \@ifxundefined [1]{%
 \@ifx{#1\undefined}
}%
\providecommand \@ifnum [1]{%
 \ifnum #1\expandafter \@firstoftwo
 \else \expandafter \@secondoftwo
 \fi
}%
\providecommand \@ifx [1]{%
 \ifx #1\expandafter \@firstoftwo
 \else \expandafter \@secondoftwo
 \fi
}%
\providecommand \natexlab [1]{#1}%
\providecommand \enquote  [1]{``#1''}%
\providecommand \bibnamefont  [1]{#1}%
\providecommand \bibfnamefont [1]{#1}%
\providecommand \citenamefont [1]{#1}%
\providecommand \href@noop [0]{\@secondoftwo}%
\providecommand \href [0]{\begingroup \@sanitize@url \@href}%
\providecommand \@href[1]{\@@startlink{#1}\@@href}%
\providecommand \@@href[1]{\endgroup#1\@@endlink}%
\providecommand \@sanitize@url [0]{\catcode `\\12\catcode `\$12\catcode
  `\&12\catcode `\#12\catcode `\^12\catcode `\_12\catcode `\%12\relax}%
\providecommand \@@startlink[1]{}%
\providecommand \@@endlink[0]{}%
\providecommand \url  [0]{\begingroup\@sanitize@url \@url }%
\providecommand \@url [1]{\endgroup\@href {#1}{\urlprefix }}%
\providecommand \urlprefix  [0]{URL }%
\providecommand \Eprint [0]{\href }%
\providecommand \doibase [0]{http://dx.doi.org/}%
\providecommand \selectlanguage [0]{\@gobble}%
\providecommand \bibinfo  [0]{\@secondoftwo}%
\providecommand \bibfield  [0]{\@secondoftwo}%
\providecommand \translation [1]{[#1]}%
\providecommand \BibitemOpen [0]{}%
\providecommand \bibitemStop [0]{}%
\providecommand \bibitemNoStop [0]{.\EOS\space}%
\providecommand \EOS [0]{\spacefactor3000\relax}%
\providecommand \BibitemShut  [1]{\csname bibitem#1\endcsname}%
\let\auto@bib@innerbib\@empty
%</preamble>
\bibitem [{\citenamefont {Amico}\ \emph {et~al.}(2008)\citenamefont {Amico},
  \citenamefont {Fazio}, \citenamefont {Osterloh},\ and\ \citenamefont
  {Vedral}}]{Amico2008}%
  \BibitemOpen
  \bibfield  {author} {\bibinfo {author} {\bibfnamefont {L.}~\bibnamefont
  {Amico}}, \bibinfo {author} {\bibfnamefont {R.}~\bibnamefont {Fazio}},
  \bibinfo {author} {\bibfnamefont {A.}~\bibnamefont {Osterloh}}, \ and\
  \bibinfo {author} {\bibfnamefont {V.}~\bibnamefont {Vedral}},\ }\href
  {\doibase 10.1103/RevModPhys.80.517} {\bibfield  {journal} {\bibinfo
  {journal} {Rev. Mod. Phys.}\ }\textbf {\bibinfo {volume} {80}},\ \bibinfo
  {pages} {517} (\bibinfo {year} {2008})}\BibitemShut {NoStop}%
\bibitem [{\citenamefont {Fradkin}(2013)}]{fradkinbook}%
  \BibitemOpen
  \bibfield  {author} {\bibinfo {author} {\bibfnamefont {E.}~\bibnamefont
  {Fradkin}},\ }\href@noop {} {\emph {\bibinfo {title} {{Field Theories of
  Condensed Matter Systems}}}}\ (\bibinfo  {publisher} {Cambridge University
  Press},\ \bibinfo {year} {2013})\BibitemShut {NoStop}%
\bibitem [{\citenamefont {Eisert}\ \emph {et~al.}(2010)\citenamefont {Eisert},
  \citenamefont {Cramer},\ and\ \citenamefont {Plenio}}]{Eisert2010}%
  \BibitemOpen
  \bibfield  {author} {\bibinfo {author} {\bibfnamefont {J.}~\bibnamefont
  {Eisert}}, \bibinfo {author} {\bibfnamefont {M.}~\bibnamefont {Cramer}}, \
  and\ \bibinfo {author} {\bibfnamefont {M.~B.}\ \bibnamefont {Plenio}},\
  }\href {\doibase 10.1103/RevModPhys.82.277} {\bibfield  {journal} {\bibinfo
  {journal} {Rev. Mod. Phys.}\ }\textbf {\bibinfo {volume} {82}},\ \bibinfo
  {pages} {277} (\bibinfo {year} {2010})}\BibitemShut {NoStop}%
\bibitem [{\citenamefont {Laflorencie}(2016)}]{Laflorencie2016}%
  \BibitemOpen
  \bibfield  {author} {\bibinfo {author} {\bibfnamefont {N.}~\bibnamefont
  {Laflorencie}},\ }\href {https://doi.org/10.1016/j.physrep.2016.06.008}
  {\bibfield  {journal} {\bibinfo  {journal} {Phys. Rep.}\ }\textbf {\bibinfo
  {volume} {646}},\ \bibinfo {pages} {1} (\bibinfo {year} {2016})}\BibitemShut
  {NoStop}%
\bibitem [{\citenamefont {Holzhey}\ \emph {et~al.}(1994)\citenamefont
  {Holzhey}, \citenamefont {Larsen},\ and\ \citenamefont
  {Wilczek}}]{Holzhey_1994}%
  \BibitemOpen
  \bibfield  {author} {\bibinfo {author} {\bibfnamefont {C.}~\bibnamefont
  {Holzhey}}, \bibinfo {author} {\bibfnamefont {F.}~\bibnamefont {Larsen}}, \
  and\ \bibinfo {author} {\bibfnamefont {F.}~\bibnamefont {Wilczek}},\ }\href
  {\doibase 10.1016/0550-3213(94)90402-2} {\bibfield  {journal} {\bibinfo
  {journal} {Nuclear Physics B}\ }\textbf {\bibinfo {volume} {424}},\ \bibinfo
  {pages} {443} (\bibinfo {year} {1994})}\BibitemShut {NoStop}%
\bibitem [{\citenamefont {Calabrese}\ and\ \citenamefont
  {Cardy}(2004)}]{Calabrese2004}%
  \BibitemOpen
  \bibfield  {author} {\bibinfo {author} {\bibfnamefont {P.}~\bibnamefont
  {Calabrese}}\ and\ \bibinfo {author} {\bibfnamefont {J.}~\bibnamefont
  {Cardy}},\ }\href {http://stacks.iop.org/1742-5468/2004/i=06/a=P06002}
  {\bibfield  {journal} {\bibinfo  {journal} {Journal of Statistical Mechanics:
  Theory and Experiment}\ }\textbf {\bibinfo {volume} {2004}},\ \bibinfo
  {pages} {P06002} (\bibinfo {year} {2004})}\BibitemShut {NoStop}%
\bibitem [{\citenamefont {Calabrese}\ and\ \citenamefont
  {Cardy}(2009)}]{Calabrese_2009}%
  \BibitemOpen
  \bibfield  {author} {\bibinfo {author} {\bibfnamefont {P.}~\bibnamefont
  {Calabrese}}\ and\ \bibinfo {author} {\bibfnamefont {J.}~\bibnamefont
  {Cardy}},\ }\href {\doibase 10.1088/1751-8113/42/50/504005} {\bibfield
  {journal} {\bibinfo  {journal} {J. Phys. A: Math. Theor.}\ }\textbf {\bibinfo
  {volume} {42}},\ \bibinfo {pages} {504005} (\bibinfo {year}
  {2009})}\BibitemShut {NoStop}%
\bibitem [{\citenamefont {Kitaev}\ and\ \citenamefont
  {Preskill}(2006)}]{Kitaev_2006}%
  \BibitemOpen
  \bibfield  {author} {\bibinfo {author} {\bibfnamefont {A.}~\bibnamefont
  {Kitaev}}\ and\ \bibinfo {author} {\bibfnamefont {J.}~\bibnamefont
  {Preskill}},\ }\href {\doibase 10.1103/PhysRevLett.96.110404} {\bibfield
  {journal} {\bibinfo  {journal} {Phys. Rev. Lett.}\ }\textbf {\bibinfo
  {volume} {96}},\ \bibinfo {pages} {110404} (\bibinfo {year}
  {2006})}\BibitemShut {NoStop}%
\bibitem [{\citenamefont {{Levin}}\ and\ \citenamefont
  {{Wen}}(2006)}]{2006PhRvL..96k0405L}%
  \BibitemOpen
  \bibfield  {author} {\bibinfo {author} {\bibfnamefont {M.}~\bibnamefont
  {{Levin}}}\ and\ \bibinfo {author} {\bibfnamefont {X.-G.}\ \bibnamefont
  {{Wen}}},\ }\href {\doibase 10.1103/PhysRevLett.96.110405} {\bibfield
  {journal} {\bibinfo  {journal} {\prl}\ }\textbf {\bibinfo {volume} {96}},\
  \bibinfo {eid} {110405} (\bibinfo {year} {2006})}\BibitemShut {NoStop}%
\bibitem [{\citenamefont {Metlitski}\ and\ \citenamefont
  {Grover}(2011)}]{Metlitski2011}%
  \BibitemOpen
  \bibfield  {author} {\bibinfo {author} {\bibfnamefont {M.~A.}\ \bibnamefont
  {Metlitski}}\ and\ \bibinfo {author} {\bibfnamefont {T.}~\bibnamefont
  {Grover}},\ }\href@noop {} {\  (\bibinfo {year} {2011})},\ \Eprint
  {http://arxiv.org/abs/1112.5166} {arXiv:1112.5166 [cond-mat.str-el]}
  \BibitemShut {NoStop}%
\bibitem [{\citenamefont {Schollw{\"{o}}ck}(2005)}]{Schollwock2005}%
  \BibitemOpen
  \bibfield  {author} {\bibinfo {author} {\bibfnamefont {U.}~\bibnamefont
  {Schollw{\"{o}}ck}},\ }\href {\doibase 10.1103/RevModPhys.77.259} {\bibfield
  {journal} {\bibinfo  {journal} {Rev. Mod. Phys.}\ }\textbf {\bibinfo {volume}
  {77}},\ \bibinfo {pages} {259} (\bibinfo {year} {2005})}\BibitemShut
  {NoStop}%
\bibitem [{\citenamefont {Moura~Alves}\ and\ \citenamefont
  {Jaksch}(2004)}]{Moura_Alves_2004}%
  \BibitemOpen
  \bibfield  {author} {\bibinfo {author} {\bibfnamefont {C.}~\bibnamefont
  {Moura~Alves}}\ and\ \bibinfo {author} {\bibfnamefont {D.}~\bibnamefont
  {Jaksch}},\ }\href {\doibase 10.1103/PhysRevLett.93.110501} {\bibfield
  {journal} {\bibinfo  {journal} {Phys. Rev. Lett.}\ }\textbf {\bibinfo
  {volume} {93}},\ \bibinfo {pages} {110501} (\bibinfo {year}
  {2004})}\BibitemShut {NoStop}%
\bibitem [{\citenamefont {Daley}\ \emph {et~al.}(2012)\citenamefont {Daley},
  \citenamefont {Pichler}, \citenamefont {Schachenmayer},\ and\ \citenamefont
  {Zoller}}]{Daley_2012}%
  \BibitemOpen
  \bibfield  {author} {\bibinfo {author} {\bibfnamefont {A.~J.}\ \bibnamefont
  {Daley}}, \bibinfo {author} {\bibfnamefont {H.}~\bibnamefont {Pichler}},
  \bibinfo {author} {\bibfnamefont {J.}~\bibnamefont {Schachenmayer}}, \ and\
  \bibinfo {author} {\bibfnamefont {P.}~\bibnamefont {Zoller}},\ }\href
  {\doibase 10.1103/PhysRevLett.109.020505} {\bibfield  {journal} {\bibinfo
  {journal} {Phys. Rev. Lett.}\ }\textbf {\bibinfo {volume} {109}},\ \bibinfo
  {pages} {020505} (\bibinfo {year} {2012})}\BibitemShut {NoStop}%
\bibitem [{\citenamefont {Alba}(2017)}]{Alba:2017aa}%
  \BibitemOpen
  \bibfield  {author} {\bibinfo {author} {\bibfnamefont {V.}~\bibnamefont
  {Alba}},\ }\href@noop {} {\bibfield  {journal} {\bibinfo  {journal} {Phys.
  Rev. E}\ }\textbf {\bibinfo {volume} {95}},\ \bibinfo {pages} {062132}
  (\bibinfo {year} {2017})}\BibitemShut {NoStop}%
\bibitem [{\citenamefont {Elben}\ \emph {et~al.}(2018)\citenamefont {Elben},
  \citenamefont {Vermersch}, \citenamefont {Dalmonte}, \citenamefont {Cirac},\
  and\ \citenamefont {Zoller}}]{Elben_2018}%
  \BibitemOpen
  \bibfield  {author} {\bibinfo {author} {\bibfnamefont {A.}~\bibnamefont
  {Elben}}, \bibinfo {author} {\bibfnamefont {B.}~\bibnamefont {Vermersch}},
  \bibinfo {author} {\bibfnamefont {M.}~\bibnamefont {Dalmonte}}, \bibinfo
  {author} {\bibfnamefont {J.~I.}\ \bibnamefont {Cirac}}, \ and\ \bibinfo
  {author} {\bibfnamefont {P.}~\bibnamefont {Zoller}},\ }\href {\doibase
  10.1103/PhysRevLett.120.050406} {\bibfield  {journal} {\bibinfo  {journal}
  {Phys. Rev. Lett.}\ }\textbf {\bibinfo {volume} {120}},\ \bibinfo {pages}
  {050406} (\bibinfo {year} {2018})}\BibitemShut {NoStop}%
\bibitem [{\citenamefont {Vermersch}\ \emph {et~al.}(2018)\citenamefont
  {Vermersch}, \citenamefont {Elben}, \citenamefont {Dalmonte}, \citenamefont
  {Cirac},\ and\ \citenamefont {Zoller}}]{Vermersch_2018}%
  \BibitemOpen
  \bibfield  {author} {\bibinfo {author} {\bibfnamefont {B.}~\bibnamefont
  {Vermersch}}, \bibinfo {author} {\bibfnamefont {A.}~\bibnamefont {Elben}},
  \bibinfo {author} {\bibfnamefont {M.}~\bibnamefont {Dalmonte}}, \bibinfo
  {author} {\bibfnamefont {J.~I.}\ \bibnamefont {Cirac}}, \ and\ \bibinfo
  {author} {\bibfnamefont {P.}~\bibnamefont {Zoller}},\ }\href {\doibase
  10.1103/PhysRevA.97.023604} {\bibfield  {journal} {\bibinfo  {journal} {Phys.
  Rev. A}\ }\textbf {\bibinfo {volume} {97}},\ \bibinfo {pages} {023604}
  (\bibinfo {year} {2018})}\BibitemShut {NoStop}%
\bibitem [{\citenamefont {Islam}\ \emph {et~al.}(2015)\citenamefont {Islam},
  \citenamefont {Ma}, \citenamefont {Preiss}, \citenamefont {Tai},
  \citenamefont {Lukin}, \citenamefont {Rispoli},\ and\ \citenamefont
  {Greiner}}]{Islam:2015aa}%
  \BibitemOpen
  \bibfield  {author} {\bibinfo {author} {\bibfnamefont {R.}~\bibnamefont
  {Islam}}, \bibinfo {author} {\bibfnamefont {R.}~\bibnamefont {Ma}}, \bibinfo
  {author} {\bibfnamefont {P.~M.}\ \bibnamefont {Preiss}}, \bibinfo {author}
  {\bibfnamefont {M.~E.}\ \bibnamefont {Tai}}, \bibinfo {author} {\bibfnamefont
  {A.}~\bibnamefont {Lukin}}, \bibinfo {author} {\bibfnamefont
  {M.}~\bibnamefont {Rispoli}}, \ and\ \bibinfo {author} {\bibfnamefont
  {M.}~\bibnamefont {Greiner}},\ }\href@noop {} {\bibfield  {journal} {\bibinfo
   {journal} {Nature}\ }\textbf {\bibinfo {volume} {528}},\ \bibinfo {pages}
  {77} (\bibinfo {year} {2015})}\BibitemShut {NoStop}%
\bibitem [{\citenamefont {Brydges}\ \emph {et~al.}(2018)\citenamefont
  {Brydges}, \citenamefont {Elben}, \citenamefont {Jurcevic}, \citenamefont
  {Vermersch}, \citenamefont {Maier}, \citenamefont {Lanyon}, \citenamefont
  {Zoller}, \citenamefont {Blatt},\ and\ \citenamefont
  {Roos}}]{Brydges:2018aa}%
  \BibitemOpen
  \bibfield  {author} {\bibinfo {author} {\bibfnamefont {T.}~\bibnamefont
  {Brydges}}, \bibinfo {author} {\bibfnamefont {A.}~\bibnamefont {Elben}},
  \bibinfo {author} {\bibfnamefont {P.}~\bibnamefont {Jurcevic}}, \bibinfo
  {author} {\bibfnamefont {B.}~\bibnamefont {Vermersch}}, \bibinfo {author}
  {\bibfnamefont {C.}~\bibnamefont {Maier}}, \bibinfo {author} {\bibfnamefont
  {B.~P.}\ \bibnamefont {Lanyon}}, \bibinfo {author} {\bibfnamefont
  {P.}~\bibnamefont {Zoller}}, \bibinfo {author} {\bibfnamefont
  {R.}~\bibnamefont {Blatt}}, \ and\ \bibinfo {author} {\bibfnamefont {C.~F.}\
  \bibnamefont {Roos}},\ }\href {https://arxiv.org/pdf/1806.05747} {\
  (\bibinfo {year} {2018})},\ \Eprint {http://arxiv.org/abs/arXiv:1806.05747}
  {arXiv:1806.05747} \BibitemShut {NoStop}%
\bibitem [{\citenamefont {Kaul}\ \emph {et~al.}(2013)\citenamefont {Kaul},
  \citenamefont {Melko},\ and\ \citenamefont {Sandvik}}]{Kaul:2013aa}%
  \BibitemOpen
  \bibfield  {author} {\bibinfo {author} {\bibfnamefont {R.~K.}\ \bibnamefont
  {Kaul}}, \bibinfo {author} {\bibfnamefont {R.~G.}\ \bibnamefont {Melko}}, \
  and\ \bibinfo {author} {\bibfnamefont {A.~W.}\ \bibnamefont {Sandvik}},\
  }\href@noop {} {\bibfield  {journal} {\bibinfo  {journal} {Annu. Rev. Con.
  Mat. Phys.}\ }\textbf {\bibinfo {volume} {4}},\ \bibinfo {pages} {179}
  (\bibinfo {year} {2013})}\BibitemShut {NoStop}%
\bibitem [{\citenamefont {Bisognano}\ and\ \citenamefont
  {Wichmann}(1975)}]{bisognano1975duality}%
  \BibitemOpen
  \bibfield  {author} {\bibinfo {author} {\bibfnamefont {J.~J.}\ \bibnamefont
  {Bisognano}}\ and\ \bibinfo {author} {\bibfnamefont {E.~H.}\ \bibnamefont
  {Wichmann}},\ }\href {http://dx.doi.org/10.1063/1.522605} {\bibfield
  {journal} {\bibinfo  {journal} {J. Math. Phys.}\ }\textbf {\bibinfo {volume}
  {16}},\ \bibinfo {pages} {985} (\bibinfo {year} {1975})}\BibitemShut
  {NoStop}%
\bibitem [{\citenamefont {Bisognano}\ and\ \citenamefont
  {Wichmann}(1976)}]{bisognano1976duality}%
  \BibitemOpen
  \bibfield  {author} {\bibinfo {author} {\bibfnamefont {J.~J.}\ \bibnamefont
  {Bisognano}}\ and\ \bibinfo {author} {\bibfnamefont {E.~H.}\ \bibnamefont
  {Wichmann}},\ }\href {https://doi.org/10.1063/1.522898} {\bibfield  {journal}
  {\bibinfo  {journal} {J. Math. Phys.}\ }\textbf {\bibinfo {volume} {17}},\
  \bibinfo {pages} {303} (\bibinfo {year} {1976})}\BibitemShut {NoStop}%
\bibitem [{\citenamefont {Dalmonte}\ \emph {et~al.}(2018)\citenamefont
  {Dalmonte}, \citenamefont {Vermersch},\ and\ \citenamefont
  {Zoller}}]{Dalmonte:2017aa}%
  \BibitemOpen
  \bibfield  {author} {\bibinfo {author} {\bibfnamefont {M.}~\bibnamefont
  {Dalmonte}}, \bibinfo {author} {\bibfnamefont {B.}~\bibnamefont {Vermersch}},
  \ and\ \bibinfo {author} {\bibfnamefont {P.}~\bibnamefont {Zoller}},\ }\href
  {https://arxiv.org/pdf/1707.04455} {\bibfield  {journal} {\bibinfo  {journal}
  {Nat. Phys.}\ }\textbf {\bibinfo {volume} {14}},\ \bibinfo {pages} {827}
  (\bibinfo {year} {2018})}\BibitemShut {NoStop}%
\bibitem [{\citenamefont {Giudici}\ \emph {et~al.}(2018)\citenamefont
  {Giudici}, \citenamefont {Mendes-Santos}, \citenamefont {Calabrese},\ and\
  \citenamefont {Dalmonte}}]{Giuliano2018}%
  \BibitemOpen
  \bibfield  {author} {\bibinfo {author} {\bibfnamefont {G.}~\bibnamefont
  {Giudici}}, \bibinfo {author} {\bibfnamefont {T.}~\bibnamefont
  {Mendes-Santos}}, \bibinfo {author} {\bibfnamefont {P.}~\bibnamefont
  {Calabrese}}, \ and\ \bibinfo {author} {\bibfnamefont {M.}~\bibnamefont
  {Dalmonte}},\ }\href {\doibase 10.1103/PhysRevB.98.134403} {\bibfield
  {journal} {\bibinfo  {journal} {Phys. Rev. B}\ }\textbf {\bibinfo {volume}
  {98}},\ \bibinfo {pages} {134403} (\bibinfo {year} {2018})}\BibitemShut
  {NoStop}%
\bibitem [{\citenamefont {Witten}(2018)}]{Witten:2018aa}%
  \BibitemOpen
  \bibfield  {author} {\bibinfo {author} {\bibfnamefont {E.}~\bibnamefont
  {Witten}},\ }\href {https://arxiv.org/pdf/1803.04993} {\bibfield  {journal}
  {\bibinfo  {journal} {Rev. Mod. Phys.}\ }\textbf {\bibinfo {volume} {90}},\
  \bibinfo {pages} {45003} (\bibinfo {year} {2018})},\ \Eprint
  {http://arxiv.org/abs/1803.04993} {1803.04993} \BibitemShut {NoStop}%
\bibitem [{\citenamefont {Abanin}\ and\ \citenamefont
  {Demler}(2012)}]{Abanin:2012aa}%
  \BibitemOpen
  \bibfield  {author} {\bibinfo {author} {\bibfnamefont {D.~A.}\ \bibnamefont
  {Abanin}}\ and\ \bibinfo {author} {\bibfnamefont {E.}~\bibnamefont
  {Demler}},\ }\href@noop {} {\bibfield  {journal} {\bibinfo  {journal} {Phys.
  Rev. Lett.}\ }\textbf {\bibinfo {volume} {109}},\ \bibinfo {pages} {020504}
  (\bibinfo {year} {2012})}\BibitemShut {NoStop}%
\bibitem [{\citenamefont {Pichler}\ \emph {et~al.}(2016)\citenamefont
  {Pichler}, \citenamefont {Zhu}, \citenamefont {Seif}, \citenamefont
  {Zoller},\ and\ \citenamefont {Hafezi}}]{Pichler:2016aa}%
  \BibitemOpen
  \bibfield  {author} {\bibinfo {author} {\bibfnamefont {H.}~\bibnamefont
  {Pichler}}, \bibinfo {author} {\bibfnamefont {G.}~\bibnamefont {Zhu}},
  \bibinfo {author} {\bibfnamefont {A.}~\bibnamefont {Seif}}, \bibinfo {author}
  {\bibfnamefont {P.}~\bibnamefont {Zoller}}, \ and\ \bibinfo {author}
  {\bibfnamefont {M.}~\bibnamefont {Hafezi}},\ }\href@noop {} {\bibfield
  {journal} {\bibinfo  {journal} {Phys. Rev. X}\ }\textbf {\bibinfo {volume}
  {6}},\ \bibinfo {pages} {041033} (\bibinfo {year} {2016})}\BibitemShut
  {NoStop}%
\bibitem [{\citenamefont {Metlitski}\ \emph {et~al.}(2009)\citenamefont
  {Metlitski}, \citenamefont {Fuertes},\ and\ \citenamefont
  {Sachdev}}]{Sachdev2009}%
  \BibitemOpen
  \bibfield  {author} {\bibinfo {author} {\bibfnamefont {M.~A.}\ \bibnamefont
  {Metlitski}}, \bibinfo {author} {\bibfnamefont {C.~A.}\ \bibnamefont
  {Fuertes}}, \ and\ \bibinfo {author} {\bibfnamefont {S.}~\bibnamefont
  {Sachdev}},\ }\href {\doibase 10.1103/PhysRevB.80.115122} {\bibfield
  {journal} {\bibinfo  {journal} {Phys. Rev. B}\ }\textbf {\bibinfo {volume}
  {80}},\ \bibinfo {pages} {115122} (\bibinfo {year} {2009})}\BibitemShut
  {NoStop}%
\bibitem [{\citenamefont {Unruh}(1976)}]{Unruh:1976aa}%
  \BibitemOpen
  \bibfield  {author} {\bibinfo {author} {\bibfnamefont {W.~G.}\ \bibnamefont
  {Unruh}},\ }\href@noop {} {\bibfield  {journal} {\bibinfo  {journal} {Phys.
  Rev. D}\ }\textbf {\bibinfo {volume} {14}},\ \bibinfo {pages} {870} (\bibinfo
  {year} {1976})}\BibitemShut {NoStop}%
\bibitem [{\citenamefont {Hislop}\ and\ \citenamefont
  {Longo}(1982)}]{hislop1982}%
  \BibitemOpen
  \bibfield  {author} {\bibinfo {author} {\bibfnamefont {P.~D.}\ \bibnamefont
  {Hislop}}\ and\ \bibinfo {author} {\bibfnamefont {R.}~\bibnamefont {Longo}},\
  }\href {https://projecteuclid.org:443/euclid.cmp/1103921046} {\bibfield
  {journal} {\bibinfo  {journal} {Comm. Math. Phys.}\ }\textbf {\bibinfo
  {volume} {84}},\ \bibinfo {pages} {71} (\bibinfo {year} {1982})}\BibitemShut
  {NoStop}%
\bibitem [{\citenamefont {Cardy}\ and\ \citenamefont
  {Tonni}(2016)}]{Cardy:2016aa}%
  \BibitemOpen
  \bibfield  {author} {\bibinfo {author} {\bibfnamefont {J.}~\bibnamefont
  {Cardy}}\ and\ \bibinfo {author} {\bibfnamefont {E.}~\bibnamefont {Tonni}},\
  }\href {https://arxiv.org/abs/1608.01283} {\bibfield  {journal} {\bibinfo
  {journal} {J. Stat. Mech.}\ ,\ \bibinfo {pages} {123103}} (\bibinfo {year}
  {2016})}\BibitemShut {NoStop}%
\bibitem [{\citenamefont {Casini}\ \emph {et~al.}(2011)\citenamefont {Casini},
  \citenamefont {Huerta},\ and\ \citenamefont {Myers}}]{Casini:2011aa}%
  \BibitemOpen
  \bibfield  {author} {\bibinfo {author} {\bibfnamefont {H.}~\bibnamefont
  {Casini}}, \bibinfo {author} {\bibfnamefont {M.}~\bibnamefont {Huerta}}, \
  and\ \bibinfo {author} {\bibfnamefont {R.~C.}\ \bibnamefont {Myers}},\ }\href
  {https://arxiv.org/abs/1102.0440} {\bibfield  {journal} {\bibinfo  {journal}
  {JHEP}\ }\textbf {\bibinfo {volume} {1105}},\ \bibinfo {pages} {036}
  (\bibinfo {year} {2011})}\BibitemShut {NoStop}%
\bibitem [{sup()}]{supmat}%
  \BibitemOpen
  \href@noop {} {\bibinfo  {journal} {See Supplemental material which includes
  Refs.
  \cite{Bauer2011,Pereyra2008,Wang2001,Melko2013,Schollwock2006,rajabpour2019}
  for (i) a discussion of the von Neumann entropy in 1D systems with open
  boundary condition, (ii) further information about the entanglement
  Hamiltonian in two-dimensions, and (iii) the details of the quantum Monte
  Carlo simulations}\ }\BibitemShut {NoStop}%
\bibitem [{Note1()}]{Note1}%
  \BibitemOpen
\bibfield  {journal} {  }\bibinfo {note} {We note that, while in principle the
  entropy is nothing but the expectation value of the EH, the normalization
  factor $Z_{EH}$, which is not universal, makes this approach hardly
  applicable.}\BibitemShut {Stop}%
\bibitem [{\citenamefont {Peschel}\ \emph {et~al.}(1999)\citenamefont
  {Peschel}, \citenamefont {Kaulke},\ and\ \citenamefont
  {Legeza}}]{Peschel_1999}%
  \BibitemOpen
  \bibfield  {author} {\bibinfo {author} {\bibfnamefont {I.}~\bibnamefont
  {Peschel}}, \bibinfo {author} {\bibfnamefont {M.}~\bibnamefont {Kaulke}}, \
  and\ \bibinfo {author} {\bibfnamefont {{\"O}.}~\bibnamefont {Legeza}},\
  }\href {\doibase
  10.1002/(sici)1521-3889(199902)8:2<153::aid-andp153>3.0.co;2-n} {\bibfield
  {journal} {\bibinfo  {journal} {Annalen der Physik}\ }\textbf {\bibinfo
  {volume} {8}},\ \bibinfo {pages} {153} (\bibinfo {year} {1999})}\BibitemShut
  {NoStop}%
\bibitem [{\citenamefont {Peschel}\ and\ \citenamefont
  {Eisler}(2009)}]{peschel2009}%
  \BibitemOpen
  \bibfield  {author} {\bibinfo {author} {\bibfnamefont {I.}~\bibnamefont
  {Peschel}}\ and\ \bibinfo {author} {\bibfnamefont {V.}~\bibnamefont
  {Eisler}},\ }\href {http://stacks.iop.org/1751-8121/42/i=50/a=504003}
  {\bibfield  {journal} {\bibinfo  {journal} {J. Phys. A: Math. Theor.}\
  }\textbf {\bibinfo {volume} {42}},\ \bibinfo {pages} {504003} (\bibinfo
  {year} {2009})}\BibitemShut {NoStop}%
\bibitem [{\citenamefont {Itoyama}\ and\ \citenamefont
  {Thacker}(1987)}]{Itoyama:1987aa}%
  \BibitemOpen
  \bibfield  {author} {\bibinfo {author} {\bibfnamefont {H.}~\bibnamefont
  {Itoyama}}\ and\ \bibinfo {author} {\bibfnamefont {H.~B.}\ \bibnamefont
  {Thacker}},\ }\href {https://doi.org/10.1103/PhysRevLett.58.1395} {\bibfield
  {journal} {\bibinfo  {journal} {Phys. Rev. Lett.}\ }\textbf {\bibinfo
  {volume} {58}},\ \bibinfo {pages} {1395} (\bibinfo {year}
  {1987})}\BibitemShut {NoStop}%
\bibitem [{\citenamefont {Nienhuis}\ \emph {et~al.}(2009)\citenamefont
  {Nienhuis}, \citenamefont {Campostrini},\ and\ \citenamefont
  {Calabrese}}]{Nienhuis_2009}%
  \BibitemOpen
  \bibfield  {author} {\bibinfo {author} {\bibfnamefont {B.}~\bibnamefont
  {Nienhuis}}, \bibinfo {author} {\bibfnamefont {M.}~\bibnamefont
  {Campostrini}}, \ and\ \bibinfo {author} {\bibfnamefont {P.}~\bibnamefont
  {Calabrese}},\ }\href {\doibase 10.1088/1742-5468/2009/02/p02063} {\bibfield
  {journal} {\bibinfo  {journal} {Journal of Statistical Mechanics: Theory and
  Experiment}\ }\textbf {\bibinfo {volume} {2009}},\ \bibinfo {pages} {P02063}
  (\bibinfo {year} {2009})}\BibitemShut {NoStop}%
\bibitem [{\citenamefont {Kim}\ \emph {et~al.}(2016)\citenamefont {Kim},
  \citenamefont {Katsura}, \citenamefont {Trivedi},\ and\ \citenamefont
  {Han}}]{Kim_2016}%
  \BibitemOpen
  \bibfield  {author} {\bibinfo {author} {\bibfnamefont {P.}~\bibnamefont
  {Kim}}, \bibinfo {author} {\bibfnamefont {H.}~\bibnamefont {Katsura}},
  \bibinfo {author} {\bibfnamefont {N.}~\bibnamefont {Trivedi}}, \ and\
  \bibinfo {author} {\bibfnamefont {J.~H.}\ \bibnamefont {Han}},\ }\href
  {\doibase 10.1103/physrevb.94.195110} {\bibfield  {journal} {\bibinfo
  {journal} {Phys. Rev. B}\ }\textbf {\bibinfo {volume} {94}},\ \bibinfo
  {pages} {195110} (\bibinfo {year} {2016})}\BibitemShut {NoStop}%
\bibitem [{\citenamefont {Eisler}\ and\ \citenamefont
  {Peschel}(2017)}]{Eisler:2017aa}%
  \BibitemOpen
  \bibfield  {author} {\bibinfo {author} {\bibfnamefont {V.}~\bibnamefont
  {Eisler}}\ and\ \bibinfo {author} {\bibfnamefont {I.}~\bibnamefont
  {Peschel}},\ }\href {https://arxiv.org/abs/1703.08126} {\bibfield  {journal}
  {\bibinfo  {journal} {J. Phys. A: Math. Theor.}\ }\textbf {\bibinfo {volume}
  {50}},\ \bibinfo {pages} {284003} (\bibinfo {year} {2017})}\BibitemShut
  {NoStop}%
\bibitem [{\citenamefont {Eisler}\ and\ \citenamefont
  {Peschel}(2018)}]{Peschel2018}%
  \BibitemOpen
  \bibfield  {author} {\bibinfo {author} {\bibfnamefont {V.}~\bibnamefont
  {Eisler}}\ and\ \bibinfo {author} {\bibfnamefont {I.}~\bibnamefont
  {Peschel}},\ }\href {http://stacks.iop.org/1742-5468/2018/i=10/a=104001}
  {\bibfield  {journal} {\bibinfo  {journal} {Journal of Statistical Mechanics:
  Theory and Experiment}\ }\textbf {\bibinfo {volume} {2018}},\ \bibinfo
  {pages} {104001} (\bibinfo {year} {2018})}\BibitemShut {NoStop}%
\bibitem [{\citenamefont {Zhu}\ \emph {et~al.}()\citenamefont {Zhu},
  \citenamefont {Huang},\ and\ \citenamefont {He}}]{1806.08060}%
  \BibitemOpen
  \bibfield  {author} {\bibinfo {author} {\bibfnamefont {W.}~\bibnamefont
  {Zhu}}, \bibinfo {author} {\bibfnamefont {Z.}~\bibnamefont {Huang}}, \ and\
  \bibinfo {author} {\bibfnamefont {Y.-C.}\ \bibnamefont {He}},\ }\href@noop {}
  {\ }\Eprint {http://arxiv.org/abs/arXiv:1806.08060} {arXiv:1806.08060}
  \BibitemShut {NoStop}%
\bibitem [{\citenamefont {Parisen~Toldin}\ and\ \citenamefont
  {Assaad}(2018)}]{Toldin:2018aa}%
  \BibitemOpen
  \bibfield  {author} {\bibinfo {author} {\bibfnamefont {F.}~\bibnamefont
  {Parisen~Toldin}}\ and\ \bibinfo {author} {\bibfnamefont {F.~F.}\
  \bibnamefont {Assaad}},\ }\href {\doibase 10.1103/PhysRevLett.121.200602}
  {\bibfield  {journal} {\bibinfo  {journal} {Phys. Rev. Lett.}\ }\textbf
  {\bibinfo {volume} {121}},\ \bibinfo {pages} {200602} (\bibinfo {year}
  {2018})}\BibitemShut {NoStop}%
\bibitem [{\citenamefont {Kosior}\ \emph {et~al.}(2018)\citenamefont {Kosior},
  \citenamefont {Lewenstein},\ and\ \citenamefont {Celi}}]{Kosior:2018aa}%
  \BibitemOpen
  \bibfield  {author} {\bibinfo {author} {\bibfnamefont {A.}~\bibnamefont
  {Kosior}}, \bibinfo {author} {\bibfnamefont {M.}~\bibnamefont {Lewenstein}},
  \ and\ \bibinfo {author} {\bibfnamefont {A.}~\bibnamefont {Celi}},\ }\href
  {\doibase 10.21468/SciPostPhys.5.6.061} {\bibfield  {journal} {\bibinfo
  {journal} {SciPost Phys.}\ }\textbf {\bibinfo {volume} {5}},\ \bibinfo
  {pages} {61} (\bibinfo {year} {2018})}\BibitemShut {NoStop}%
\bibitem [{\citenamefont {Troyer}\ \emph {et~al.}(2003)\citenamefont {Troyer},
  \citenamefont {Wessel},\ and\ \citenamefont {Alet}}]{Wessel2003}%
  \BibitemOpen
  \bibfield  {author} {\bibinfo {author} {\bibfnamefont {M.}~\bibnamefont
  {Troyer}}, \bibinfo {author} {\bibfnamefont {S.}~\bibnamefont {Wessel}}, \
  and\ \bibinfo {author} {\bibfnamefont {F.}~\bibnamefont {Alet}},\ }\href
  {\doibase 10.1103/PhysRevLett.90.120201} {\bibfield  {journal} {\bibinfo
  {journal} {Phys. Rev. Lett.}\ }\textbf {\bibinfo {volume} {90}},\ \bibinfo
  {pages} {120201} (\bibinfo {year} {2003})}\BibitemShut {NoStop}%
\bibitem [{\citenamefont {Wessel}\ \emph {et~al.}(2007)\citenamefont {Wessel},
  \citenamefont {Stoop}, \citenamefont {Gull}, \citenamefont {Trebst},\ and\
  \citenamefont {Troyer}}]{Wessel2007}%
  \BibitemOpen
  \bibfield  {author} {\bibinfo {author} {\bibfnamefont {S.}~\bibnamefont
  {Wessel}}, \bibinfo {author} {\bibfnamefont {N.}~\bibnamefont {Stoop}},
  \bibinfo {author} {\bibfnamefont {E.}~\bibnamefont {Gull}}, \bibinfo {author}
  {\bibfnamefont {S.}~\bibnamefont {Trebst}}, \ and\ \bibinfo {author}
  {\bibfnamefont {M.}~\bibnamefont {Troyer}},\ }\href
  {http://stacks.iop.org/1742-5468/2007/i=12/a=P12005} {\bibfield  {journal}
  {\bibinfo  {journal} {Journal of Statistical Mechanics: Theory and
  Experiment}\ }\textbf {\bibinfo {volume} {2007}},\ \bibinfo {pages} {P12005}
  (\bibinfo {year} {2007})}\BibitemShut {NoStop}%
\bibitem [{cod()}]{codeALPS}%
  \BibitemOpen
  \href@noop {} {\bibinfo  {journal} {A working code that generates the
  necessary input files to run with ALPS Wang Landau (qwl) \cite{Bauer2011} can
  be found in
  https://github.com/tiagomendessantos/BW-entanglement-Hamiltonian}\
  }\BibitemShut {NoStop}%
\bibitem [{\citenamefont {Osborne}()}]{Osborne:aa}%
  \BibitemOpen
\bibfield  {journal} {  }\bibfield  {author} {\bibinfo {author} {\bibfnamefont
  {T.~J.}\ \bibnamefont {Osborne}},\ }\href@noop {} {\bibinfo  {journal}
  {arXiv:cond-mat/0605194}\ }\BibitemShut {NoStop}%
\bibitem [{\citenamefont {Schrodi}\ \emph {et~al.}(2017)\citenamefont
  {Schrodi}, \citenamefont {Silvi}, \citenamefont {Tschirsich}, \citenamefont
  {Fazio},\ and\ \citenamefont {Montangero}}]{Schrodi:2017aa}%
  \BibitemOpen
\bibfield  {journal} {  }\bibfield  {author} {\bibinfo {author} {\bibfnamefont
  {F.}~\bibnamefont {Schrodi}}, \bibinfo {author} {\bibfnamefont
  {P.}~\bibnamefont {Silvi}}, \bibinfo {author} {\bibfnamefont
  {F.}~\bibnamefont {Tschirsich}}, \bibinfo {author} {\bibfnamefont
  {R.}~\bibnamefont {Fazio}}, \ and\ \bibinfo {author} {\bibfnamefont
  {S.}~\bibnamefont {Montangero}},\ }\href@noop {} {\bibfield  {journal}
  {\bibinfo  {journal} {Phys. Rev. B}\ }\textbf {\bibinfo {volume} {96}},\
  \bibinfo {pages} {094303} (\bibinfo {year} {2017})}\BibitemShut {NoStop}%
\bibitem [{\citenamefont {Calabrese}\ and\ \citenamefont
  {Lefevre}(2008)}]{calabrese2008entanglement}%
  \BibitemOpen
  \bibfield  {author} {\bibinfo {author} {\bibfnamefont {P.}~\bibnamefont
  {Calabrese}}\ and\ \bibinfo {author} {\bibfnamefont {A.}~\bibnamefont
  {Lefevre}},\ }\href@noop {} {\bibfield  {journal} {\bibinfo  {journal}
  {Physical Review A}\ }\textbf {\bibinfo {volume} {78}},\ \bibinfo {pages}
  {032329} (\bibinfo {year} {2008})}\BibitemShut {NoStop}%
\bibitem [{\citenamefont {Catani}\ \emph {et~al.}(2009)\citenamefont {Catani},
  \citenamefont {Barontini}, \citenamefont {Lamporesi}, \citenamefont
  {Rabatti}, \citenamefont {Thalhammer}, \citenamefont {Minardi}, \citenamefont
  {Stringari},\ and\ \citenamefont {Inguscio}}]{Catani:2009aa}%
  \BibitemOpen
  \bibfield  {author} {\bibinfo {author} {\bibfnamefont {J.}~\bibnamefont
  {Catani}}, \bibinfo {author} {\bibfnamefont {G.}~\bibnamefont {Barontini}},
  \bibinfo {author} {\bibfnamefont {G.}~\bibnamefont {Lamporesi}}, \bibinfo
  {author} {\bibfnamefont {F.}~\bibnamefont {Rabatti}}, \bibinfo {author}
  {\bibfnamefont {G.}~\bibnamefont {Thalhammer}}, \bibinfo {author}
  {\bibfnamefont {F.}~\bibnamefont {Minardi}}, \bibinfo {author} {\bibfnamefont
  {S.}~\bibnamefont {Stringari}}, \ and\ \bibinfo {author} {\bibfnamefont
  {M.}~\bibnamefont {Inguscio}},\ }\href@noop {} {\bibfield  {journal}
  {\bibinfo  {journal} {Phys. Rev. Lett.}\ }\textbf {\bibinfo {volume} {103}},\
  \bibinfo {pages} {140401} (\bibinfo {year} {2009})}\BibitemShut {NoStop}%
\bibitem [{\citenamefont {White}(1992)}]{White1992}%
  \BibitemOpen
  \bibfield  {author} {\bibinfo {author} {\bibfnamefont {S.~R.}\ \bibnamefont
  {White}},\ }\href {\doibase 10.1103/PhysRevLett.69.2863} {\bibfield
  {journal} {\bibinfo  {journal} {Phys. Rev. Lett.}\ }\textbf {\bibinfo
  {volume} {69}},\ \bibinfo {pages} {2863} (\bibinfo {year}
  {1992})}\BibitemShut {NoStop}%
\bibitem [{\citenamefont {Sandvik}\ and\ \citenamefont
  {Kurkij\"arvi}(1991)}]{sandvik1991}%
  \BibitemOpen
  \bibfield  {author} {\bibinfo {author} {\bibfnamefont {A.~W.}\ \bibnamefont
  {Sandvik}}\ and\ \bibinfo {author} {\bibfnamefont {J.}~\bibnamefont
  {Kurkij\"arvi}},\ }\href {\doibase 10.1103/PhysRevB.43.5950} {\bibfield
  {journal} {\bibinfo  {journal} {Phys. Rev. B}\ }\textbf {\bibinfo {volume}
  {43}},\ \bibinfo {pages} {5950} (\bibinfo {year} {1991})}\BibitemShut
  {NoStop}%
\bibitem [{\citenamefont {Sylju\aa{}sen}\ and\ \citenamefont
  {Sandvik}(2002)}]{sandvik2002}%
  \BibitemOpen
  \bibfield  {author} {\bibinfo {author} {\bibfnamefont {O.~F.}\ \bibnamefont
  {Sylju\aa{}sen}}\ and\ \bibinfo {author} {\bibfnamefont {A.~W.}\ \bibnamefont
  {Sandvik}},\ }\href {\doibase 10.1103/PhysRevE.66.046701} {\bibfield
  {journal} {\bibinfo  {journal} {Phys. Rev. E}\ }\textbf {\bibinfo {volume}
  {66}},\ \bibinfo {pages} {046701} (\bibinfo {year} {2002})}\BibitemShut
  {NoStop}%
\bibitem [{\citenamefont {Laflorencie}\ \emph {et~al.}(2006)\citenamefont
  {Laflorencie}, \citenamefont {S\o{}rensen}, \citenamefont {Chang},\ and\
  \citenamefont {Affleck}}]{Laflorencie2006}%
  \BibitemOpen
  \bibfield  {author} {\bibinfo {author} {\bibfnamefont {N.}~\bibnamefont
  {Laflorencie}}, \bibinfo {author} {\bibfnamefont {E.~S.}\ \bibnamefont
  {S\o{}rensen}}, \bibinfo {author} {\bibfnamefont {M.-S.}\ \bibnamefont
  {Chang}}, \ and\ \bibinfo {author} {\bibfnamefont {I.}~\bibnamefont
  {Affleck}},\ }\href {\doibase 10.1103/PhysRevLett.96.100603} {\bibfield
  {journal} {\bibinfo  {journal} {Phys. Rev. Lett.}\ }\textbf {\bibinfo
  {volume} {96}},\ \bibinfo {pages} {100603} (\bibinfo {year}
  {2006})}\BibitemShut {NoStop}%
\bibitem [{\citenamefont {Calabrese}\ \emph {et~al.}(2010)\citenamefont
  {Calabrese}, \citenamefont {Campostrini}, \citenamefont {Essler},\ and\
  \citenamefont {Nienhuis}}]{Calabrese2010}%
  \BibitemOpen
  \bibfield  {author} {\bibinfo {author} {\bibfnamefont {P.}~\bibnamefont
  {Calabrese}}, \bibinfo {author} {\bibfnamefont {M.}~\bibnamefont
  {Campostrini}}, \bibinfo {author} {\bibfnamefont {F.}~\bibnamefont {Essler}},
  \ and\ \bibinfo {author} {\bibfnamefont {B.}~\bibnamefont {Nienhuis}},\
  }\href {\doibase 10.1103/PhysRevLett.104.095701} {\bibfield  {journal}
  {\bibinfo  {journal} {Phys. Rev. Lett.}\ }\textbf {\bibinfo {volume} {104}},\
  \bibinfo {pages} {1} (\bibinfo {year} {2010})},\ \Eprint
  {http://arxiv.org/abs/0911.4660} {arXiv:0911.4660} \BibitemShut {NoStop}%
\bibitem [{\citenamefont {Kallin}\ \emph {et~al.}(2011)\citenamefont {Kallin},
  \citenamefont {Hastings}, \citenamefont {Melko},\ and\ \citenamefont
  {Singh}}]{Singh2011}%
  \BibitemOpen
  \bibfield  {author} {\bibinfo {author} {\bibfnamefont {A.~B.}\ \bibnamefont
  {Kallin}}, \bibinfo {author} {\bibfnamefont {M.~B.}\ \bibnamefont
  {Hastings}}, \bibinfo {author} {\bibfnamefont {R.~G.}\ \bibnamefont {Melko}},
  \ and\ \bibinfo {author} {\bibfnamefont {R.~R.~P.}\ \bibnamefont {Singh}},\
  }\href {\doibase 10.1103/PhysRevB.84.165134} {\bibfield  {journal} {\bibinfo
  {journal} {Phys. Rev. B}\ }\textbf {\bibinfo {volume} {84}},\ \bibinfo
  {pages} {165134} (\bibinfo {year} {2011})}\BibitemShut {NoStop}%
\bibitem [{\citenamefont {Sen}\ \emph {et~al.}(2015)\citenamefont {Sen},
  \citenamefont {Suwa},\ and\ \citenamefont {Sandvik}}]{sandvik2015}%
  \BibitemOpen
  \bibfield  {author} {\bibinfo {author} {\bibfnamefont {A.}~\bibnamefont
  {Sen}}, \bibinfo {author} {\bibfnamefont {H.}~\bibnamefont {Suwa}}, \ and\
  \bibinfo {author} {\bibfnamefont {A.~W.}\ \bibnamefont {Sandvik}},\ }\href
  {\doibase 10.1103/PhysRevB.92.195145} {\bibfield  {journal} {\bibinfo
  {journal} {Phys. Rev. B}\ }\textbf {\bibinfo {volume} {92}},\ \bibinfo
  {pages} {195145} (\bibinfo {year} {2015})}\BibitemShut {NoStop}%
\bibitem [{\citenamefont {Sandvik}\ and\ \citenamefont
  {Hamer}(1999)}]{sandvik1999}%
  \BibitemOpen
  \bibfield  {author} {\bibinfo {author} {\bibfnamefont {A.~W.}\ \bibnamefont
  {Sandvik}}\ and\ \bibinfo {author} {\bibfnamefont {C.~J.}\ \bibnamefont
  {Hamer}},\ }\href {\doibase 10.1103/PhysRevB.60.6588} {\bibfield  {journal}
  {\bibinfo  {journal} {Phys. Rev. B}\ }\textbf {\bibinfo {volume} {60}},\
  \bibinfo {pages} {6588} (\bibinfo {year} {1999})}\BibitemShut {NoStop}%
\bibitem [{\citenamefont {Song}\ \emph {et~al.}(2011)\citenamefont {Song},
  \citenamefont {Laflorencie}, \citenamefont {Rachel},\ and\ \citenamefont
  {Le~Hur}}]{Laflorencie2011}%
  \BibitemOpen
  \bibfield  {author} {\bibinfo {author} {\bibfnamefont {H.~F.}\ \bibnamefont
  {Song}}, \bibinfo {author} {\bibfnamefont {N.}~\bibnamefont {Laflorencie}},
  \bibinfo {author} {\bibfnamefont {S.}~\bibnamefont {Rachel}}, \ and\ \bibinfo
  {author} {\bibfnamefont {K.}~\bibnamefont {Le~Hur}},\ }\href {\doibase
  10.1103/PhysRevB.83.224410} {\bibfield  {journal} {\bibinfo  {journal} {Phys.
  Rev. B}\ }\textbf {\bibinfo {volume} {83}},\ \bibinfo {pages} {224410}
  (\bibinfo {year} {2011})}\BibitemShut {NoStop}%
\bibitem [{\citenamefont {Humeniuk}\ and\ \citenamefont
  {Roscilde}(2012)}]{Humeniuk:2012aa}%
  \BibitemOpen
  \bibfield  {author} {\bibinfo {author} {\bibfnamefont {S.}~\bibnamefont
  {Humeniuk}}\ and\ \bibinfo {author} {\bibfnamefont {T.}~\bibnamefont
  {Roscilde}},\ }\href@noop {} {\bibfield  {journal} {\bibinfo  {journal}
  {Phys. Rev. B}\ }\textbf {\bibinfo {volume} {86}} (\bibinfo {year}
  {2012})}\BibitemShut {NoStop}%
\bibitem [{\citenamefont {Kulchytskyy}\ \emph {et~al.}(2015)\citenamefont
  {Kulchytskyy}, \citenamefont {Herdman}, \citenamefont {Inglis},\ and\
  \citenamefont {Melko}}]{Melko2015}%
  \BibitemOpen
  \bibfield  {author} {\bibinfo {author} {\bibfnamefont {B.}~\bibnamefont
  {Kulchytskyy}}, \bibinfo {author} {\bibfnamefont {C.~M.}\ \bibnamefont
  {Herdman}}, \bibinfo {author} {\bibfnamefont {S.}~\bibnamefont {Inglis}}, \
  and\ \bibinfo {author} {\bibfnamefont {R.~G.}\ \bibnamefont {Melko}},\ }\href
  {\doibase 10.1103/PhysRevB.92.115146} {\bibfield  {journal} {\bibinfo
  {journal} {Phys. Rev. B}\ }\textbf {\bibinfo {volume} {92}},\ \bibinfo
  {pages} {115146} (\bibinfo {year} {2015})}\BibitemShut {NoStop}%
\bibitem [{\citenamefont {Dalmonte}\ \emph {et~al.}(2011)\citenamefont
  {Dalmonte}, \citenamefont {Ercolessi},\ and\ \citenamefont
  {Taddia}}]{Dalmonte:2011aa}%
  \BibitemOpen
  \bibfield  {author} {\bibinfo {author} {\bibfnamefont {M.}~\bibnamefont
  {Dalmonte}}, \bibinfo {author} {\bibfnamefont {E.}~\bibnamefont {Ercolessi}},
  \ and\ \bibinfo {author} {\bibfnamefont {L.}~\bibnamefont {Taddia}},\ }\href
  {https://doi.org/10.1103/PhysRevB.84.085110} {\bibfield  {journal} {\bibinfo
  {journal} {Phys. Rev. B}\ }\textbf {\bibinfo {volume} {84}},\ \bibinfo
  {pages} {085110} (\bibinfo {year} {2011})}\BibitemShut {NoStop}%
\bibitem [{\citenamefont {Sandvik}\ and\ \citenamefont
  {Scalapino}(1994)}]{sandvik1994}%
  \BibitemOpen
  \bibfield  {author} {\bibinfo {author} {\bibfnamefont {A.~W.}\ \bibnamefont
  {Sandvik}}\ and\ \bibinfo {author} {\bibfnamefont {D.~J.}\ \bibnamefont
  {Scalapino}},\ }\href {\doibase 10.1103/PhysRevLett.72.2777} {\bibfield
  {journal} {\bibinfo  {journal} {Phys. Rev. Lett.}\ }\textbf {\bibinfo
  {volume} {72}},\ \bibinfo {pages} {2777} (\bibinfo {year}
  {1994})}\BibitemShut {NoStop}%
\bibitem [{\citenamefont {Wang}\ \emph {et~al.}(2006)\citenamefont {Wang},
  \citenamefont {Beach},\ and\ \citenamefont {Sandvik}}]{sandvik2006}%
  \BibitemOpen
  \bibfield  {author} {\bibinfo {author} {\bibfnamefont {L.}~\bibnamefont
  {Wang}}, \bibinfo {author} {\bibfnamefont {K.~S.~D.}\ \bibnamefont {Beach}},
  \ and\ \bibinfo {author} {\bibfnamefont {A.~W.}\ \bibnamefont {Sandvik}},\
  }\href {\doibase 10.1103/PhysRevB.73.014431} {\bibfield  {journal} {\bibinfo
  {journal} {Phys. Rev. B}\ }\textbf {\bibinfo {volume} {73}},\ \bibinfo
  {pages} {014431} (\bibinfo {year} {2006})}\BibitemShut {NoStop}%
\bibitem [{Note2()}]{Note2}%
  \BibitemOpen
  \bibinfo {note} {We remind that this imperfect EH corresponds to the GS of a
  clean system, and is not related to the entanglement properties of disordered
  systems.}\BibitemShut {Stop}%
\bibitem [{\citenamefont {Frerot}(2017)}]{frerot2017}%
  \BibitemOpen
  \bibfield  {author} {\bibinfo {author} {\bibfnamefont {I.}~\bibnamefont
  {Frerot}},\ }\emph {\bibinfo {title} {A quantum statistical approach to
  quantum correlations in many-body systems}},\ \href@noop {} {Ph.D. thesis},\
  \bibinfo  {school} {Universite de Lyon} (\bibinfo {year} {2017})\BibitemShut
  {NoStop}%
\bibitem [{\citenamefont {Bauer}\ \emph {et~al.}(2011)\citenamefont {Bauer},
  \citenamefont {Carr}, \citenamefont {Evertz}, \citenamefont {Feiguin},
  \citenamefont {Freire}, \citenamefont {Fuchs}, \citenamefont {Gamper},
  \citenamefont {Gukelberger}, \citenamefont {Gull}, \citenamefont {Guertler},
  \citenamefont {Hehn}, \citenamefont {Igarashi}, \citenamefont {Isakov},
  \citenamefont {Koop}, \citenamefont {Ma}, \citenamefont {Mates},
  \citenamefont {Matsuo}, \citenamefont {Parcollet}, \citenamefont
  {Paw{\l}owski}, \citenamefont {Picon}, \citenamefont {Pollet}, \citenamefont
  {Santos}, \citenamefont {Scarola}, \citenamefont {Schollw{\"o}ck},
  \citenamefont {Silva}, \citenamefont {Surer}, \citenamefont {Todo},
  \citenamefont {Trebst}, \citenamefont {Troyer}, \citenamefont {Wall},
  \citenamefont {Werner},\ and\ \citenamefont {Wessel}}]{Bauer2011}%
  \BibitemOpen
  \bibfield  {author} {\bibinfo {author} {\bibfnamefont {B.}~\bibnamefont
  {Bauer}}, \bibinfo {author} {\bibfnamefont {L.~D.}\ \bibnamefont {Carr}},
  \bibinfo {author} {\bibfnamefont {H.~G.}\ \bibnamefont {Evertz}}, \bibinfo
  {author} {\bibfnamefont {A.}~\bibnamefont {Feiguin}}, \bibinfo {author}
  {\bibfnamefont {J.}~\bibnamefont {Freire}}, \bibinfo {author} {\bibfnamefont
  {S.}~\bibnamefont {Fuchs}}, \bibinfo {author} {\bibfnamefont
  {L.}~\bibnamefont {Gamper}}, \bibinfo {author} {\bibfnamefont
  {J.}~\bibnamefont {Gukelberger}}, \bibinfo {author} {\bibfnamefont
  {E.}~\bibnamefont {Gull}}, \bibinfo {author} {\bibfnamefont {S.}~\bibnamefont
  {Guertler}}, \bibinfo {author} {\bibfnamefont {A.}~\bibnamefont {Hehn}},
  \bibinfo {author} {\bibfnamefont {R.}~\bibnamefont {Igarashi}}, \bibinfo
  {author} {\bibfnamefont {S.~V.}\ \bibnamefont {Isakov}}, \bibinfo {author}
  {\bibfnamefont {D.}~\bibnamefont {Koop}}, \bibinfo {author} {\bibfnamefont
  {P.~N.}\ \bibnamefont {Ma}}, \bibinfo {author} {\bibfnamefont
  {P.}~\bibnamefont {Mates}}, \bibinfo {author} {\bibfnamefont
  {H.}~\bibnamefont {Matsuo}}, \bibinfo {author} {\bibfnamefont
  {O.}~\bibnamefont {Parcollet}}, \bibinfo {author} {\bibfnamefont
  {G.}~\bibnamefont {Paw{\l}owski}}, \bibinfo {author} {\bibfnamefont {J.~D.}\
  \bibnamefont {Picon}}, \bibinfo {author} {\bibfnamefont {L.}~\bibnamefont
  {Pollet}}, \bibinfo {author} {\bibfnamefont {E.}~\bibnamefont {Santos}},
  \bibinfo {author} {\bibfnamefont {V.~W.}\ \bibnamefont {Scarola}}, \bibinfo
  {author} {\bibfnamefont {U.}~\bibnamefont {Schollw{\"o}ck}}, \bibinfo
  {author} {\bibfnamefont {C.}~\bibnamefont {Silva}}, \bibinfo {author}
  {\bibfnamefont {B.}~\bibnamefont {Surer}}, \bibinfo {author} {\bibfnamefont
  {S.}~\bibnamefont {Todo}}, \bibinfo {author} {\bibfnamefont {S.}~\bibnamefont
  {Trebst}}, \bibinfo {author} {\bibfnamefont {M.}~\bibnamefont {Troyer}},
  \bibinfo {author} {\bibfnamefont {M.~L.}\ \bibnamefont {Wall}}, \bibinfo
  {author} {\bibfnamefont {P.}~\bibnamefont {Werner}}, \ and\ \bibinfo {author}
  {\bibfnamefont {S.}~\bibnamefont {Wessel}},\ }\href {\doibase
  10.1088/1742-5468/2011/05/p05001} {\bibfield  {journal} {\bibinfo  {journal}
  {Journal of Statistical Mechanics: Theory and Experiment}\ }\textbf {\bibinfo
  {volume} {2011}},\ \bibinfo {pages} {P05001} (\bibinfo {year}
  {2011})}\BibitemShut {NoStop}%
\bibitem [{\citenamefont {Belardinelli}\ \emph {et~al.}(2008)\citenamefont
  {Belardinelli}, \citenamefont {Manzi},\ and\ \citenamefont
  {Pereyra}}]{Pereyra2008}%
  \BibitemOpen
  \bibfield  {author} {\bibinfo {author} {\bibfnamefont {R.~E.}\ \bibnamefont
  {Belardinelli}}, \bibinfo {author} {\bibfnamefont {S.}~\bibnamefont {Manzi}},
  \ and\ \bibinfo {author} {\bibfnamefont {V.~D.}\ \bibnamefont {Pereyra}},\
  }\href {\doibase 10.1103/PhysRevE.78.067701} {\bibfield  {journal} {\bibinfo
  {journal} {Phys. Rev. E}\ }\textbf {\bibinfo {volume} {78}},\ \bibinfo
  {pages} {067701} (\bibinfo {year} {2008})}\BibitemShut {NoStop}%
\bibitem [{\citenamefont {Wang}\ and\ \citenamefont {Landau}(2001)}]{Wang2001}%
  \BibitemOpen
  \bibfield  {author} {\bibinfo {author} {\bibfnamefont {F.}~\bibnamefont
  {Wang}}\ and\ \bibinfo {author} {\bibfnamefont {D.~P.}\ \bibnamefont
  {Landau}},\ }\href {\doibase 10.1103/PhysRevLett.86.2050} {\bibfield
  {journal} {\bibinfo  {journal} {Phys. Rev. Lett.}\ }\textbf {\bibinfo
  {volume} {86}},\ \bibinfo {pages} {2050} (\bibinfo {year}
  {2001})}\BibitemShut {NoStop}%
\bibitem [{\citenamefont {Inglis}\ and\ \citenamefont
  {Melko}(2013)}]{Melko2013}%
  \BibitemOpen
  \bibfield  {author} {\bibinfo {author} {\bibfnamefont {S.}~\bibnamefont
  {Inglis}}\ and\ \bibinfo {author} {\bibfnamefont {R.~G.}\ \bibnamefont
  {Melko}},\ }\href {\doibase 10.1103/PhysRevE.87.013306} {\bibfield  {journal}
  {\bibinfo  {journal} {Phys. Rev. E}\ }\textbf {\bibinfo {volume} {87}},\
  \bibinfo {pages} {013306} (\bibinfo {year} {2013})}\BibitemShut {NoStop}%
\bibitem [{\citenamefont {Zhou}\ \emph {et~al.}(2006)\citenamefont {Zhou},
  \citenamefont {Barthel}, \citenamefont {Fj\ae{}restad},\ and\ \citenamefont
  {Schollw\"ock}}]{Schollwock2006}%
  \BibitemOpen
  \bibfield  {author} {\bibinfo {author} {\bibfnamefont {H.-Q.}\ \bibnamefont
  {Zhou}}, \bibinfo {author} {\bibfnamefont {T.}~\bibnamefont {Barthel}},
  \bibinfo {author} {\bibfnamefont {J.~O.}\ \bibnamefont {Fj\ae{}restad}}, \
  and\ \bibinfo {author} {\bibfnamefont {U.}~\bibnamefont {Schollw\"ock}},\
  }\href {\doibase 10.1103/PhysRevA.74.050305} {\bibfield  {journal} {\bibinfo
  {journal} {Phys. Rev. A}\ }\textbf {\bibinfo {volume} {74}},\ \bibinfo
  {pages} {050305} (\bibinfo {year} {2006})}\BibitemShut {NoStop}%
\bibitem [{\citenamefont {Mendes-Santos~et al.}()}]{rajabpour2019}%
  \BibitemOpen
  \bibfield  {author} {\bibinfo {author} {\bibfnamefont {T.}~\bibnamefont
  {Mendes-Santos~et al.}},\ }\href@noop {} {\bibinfo  {journal} {to be
  submitted}\ }\BibitemShut {NoStop}%
\end{thebibliography}%

\
\newpage

\widetext
\clearpage

\onecolumngrid
\begin{center}
  \textbf{\large Supplementary Material: \\ \medskip
\large Measuring many-body entanglement without wave functions}\\[.2cm]
\end{center}

\twocolumngrid

\maketitle

\twocolumngrid

\section{Entanglement entropy of one-dimensional critical systems}

\subsection{Entanglement entropy of Heisenberg and Ising models on open chains}

\begin{figure}[t]
\includegraphics[width=0.99\columnwidth]{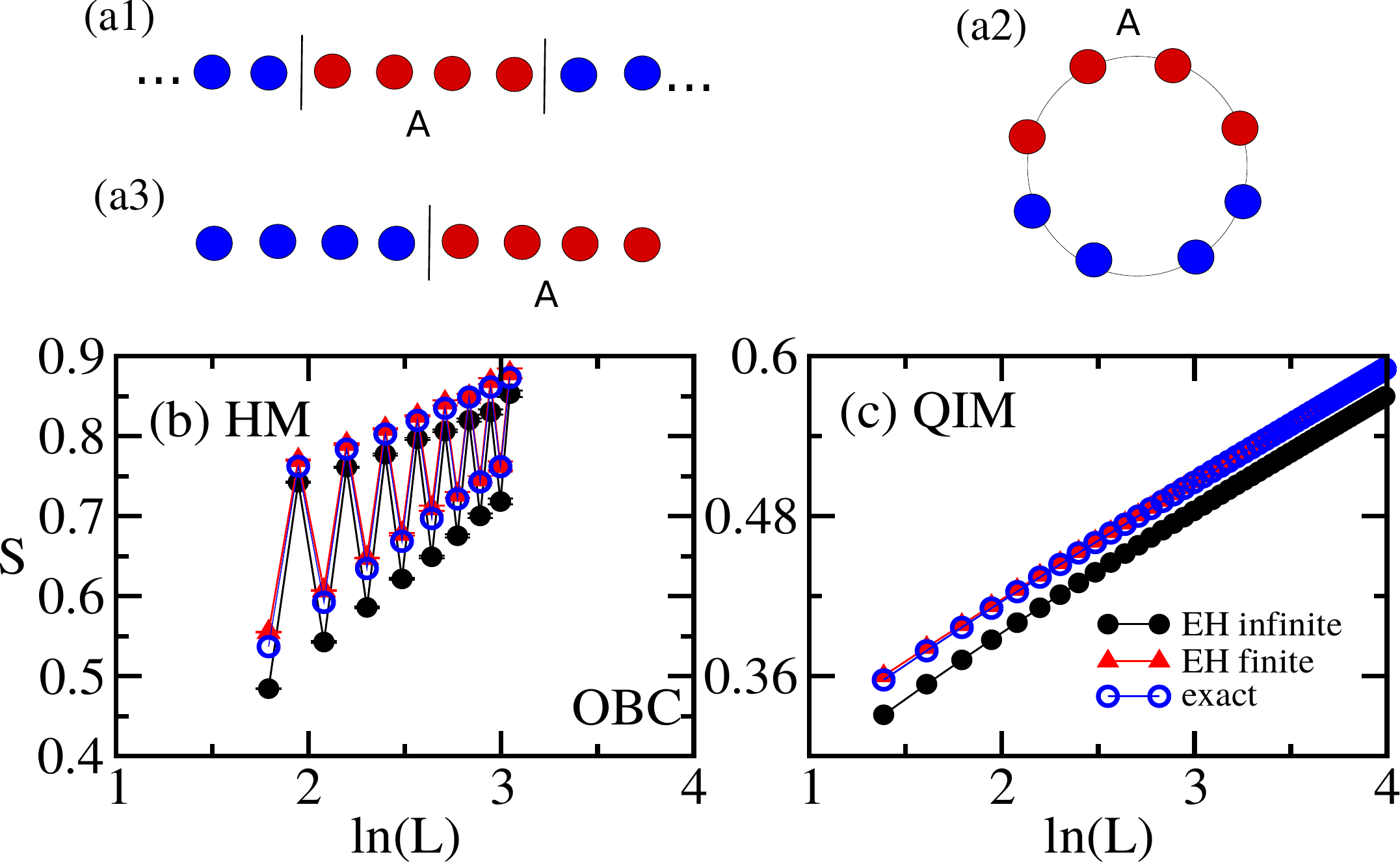}
\centering
\caption{\textit{BW-EH of one-dimensional critical systems.}
Panel (a): partitions of the one-dimensional systems that we consider:
(a1)  partition of length $L$ embedded in an infinite system (infinite PBC); (a2) half-partition of a ring (finite PBC), (a3) half-partition of an open system (finite OBC).
The BW couplings of these systems are given by the CFT generalization of the BW theorem (see text).
Panels (b-c): von Neumann entropies obtained from the BW-EH for the HM and QIM with OBC, respectively.
Error bars are smaller than the size of the symbols.}
\label{fig:1d}
\end{figure}

In this section, we provide additional details about the calculation of the von Neumann entropy (VNE) in one-dimensional critical systems.
As discussed in the main text, the VNE is related to the thermal entropy of the lattice Bisognano-Wichmann entanglement Hamiltonian (BW-EH).
For the Heisenberg model (HM) one has
\begin{align}
H_{BW} =\sum_{i} \Gamma(i) \vec{S}_{i} \vec{S}_{i+1} ,
\label{xxz}
\end{align}
and for the quantum Ising model (QIM) at the critical point
\begin{align}
H=-\sum_{i} \Gamma(i) S^z_{i} S^z_{i+1} -\sum_{i} \Gamma(i - 1/2) S^x_i.
\label{qim}
\end{align}
The low energy properties of these lattice models are described by Lorentz-invariant quantum field theories, a prerequisite for the applicability of our approach. Additionally, the emergence of conformal symmetries in these systems allows us to consider the extensions of the BW theorem to different geometrical partitions \cite{hislop1982,Cardy:2016aa}, see Fig.\ref{fig:1d} (a). For the half-partition of a ring with length $2L$, for instance, one has
\begin{align}
 \Gamma (x) = \frac{L}{ \pi} \sin \left( \frac{  \pi x }{ L } \right),
\end{align}
while for the open chain
\begin{align}
  \Gamma (x) = \frac{2L}{\pi} \sin \left(  \frac{ \pi x }{ 2L } \right).
\end{align}
The definition of the BW-EH for finite systems allows us to directly compare the BW VNE with the exact
results of the VNE obtained with exact diagonalization or the density-matrix-renormalization-group.
In the PBC case, for instance, the BW entropy  is in  perfect agreement with exact results, see Fig. 2 of the main text.
The discrepancy $\Delta S(L) = |S_{BW} - S_{exact}|$ goes to zero in the limit $L \to \infty$.
Furthermore, the corresponding central charge considering system sizes up to $L = 80$ ($L=100$)  is only
$1\%$ ($0.05\%$) away from the exact results for the HM (QIM). 
It is interesting to note that, when we directly compare the absolute value of the BW-EH entropy for the infinite PBC with the exact VNE (subsystem with size $L$ embedded in a system of size $2L$), we observe a shift~$\epsilon$.
This shift is almost equal to $\epsilon = \left( c/3 \right) \ln(\pi/2)$, as expected from the finite-size corrections given by the CFT expression of the VNE \cite{Calabrese2004,Calabrese_2009}.

For the OBC case, we observe an alternating term of the BW-EH entropy for the HM, but  not for the QIM, see Figs. \ref{fig:1d} (b) and (c).
These results is in agreement with the exact VNE.
As discussed in Ref.~\cite{Laflorencie2006,Calabrese2010}, those oscillations are universal and due to the antiferromagnetic nature of the interactions, 
not appearing in the QIM~\cite{Schollwock2006} (in the latter, the effective Fermi momentum is either 0 or $\pi$). From the CFT perspective,
the oscillations can be viewed as lattice corrections of scaling dimension $\Delta_p$: 
their decay as a function of the bipartition size is a power law whose exponent is related to $\Delta_p$ ~\cite{Calabrese2010,Dalmonte:2011aa}. 
The fact that the BW-EH faithfully reproduces not only the leading, but also the dominant subleading correction testifies its 
predictive power on generic universal quantities captured by the VNE (a CFT-specific analysis will be reported elsewhere~\cite{rajabpour2019}).
While, for instance, non-universal contributions such as additive constants in 1D shall not be immediately reproduced due to the field theoretical origin of the relation we employ, 
in all examples where a comparison to exact results is possible (essentially, 1D systems), we observe that even non-universal contributions are accurately captured: for instance, $\Delta S(L)$  
goes to zero in the limit $L \to \infty$ both in the OBC and PBC cases. We attribute this to the fact that the BW-EH is actually able to reproduce a ``partition function'' 
whose corresponding Hamiltonian has the correct density of states, and whose generic correlation functions are correct~\cite{Giuliano2018}. 
In case only the first element was true, and, for instance, the overall scaling correction was wrong, one would have generically expected incorrect correlation functions. 
From a methodological viewpoint, this implies that our method may be used to check convergence of tensor network states in conformal phases, especially for large values of the central charge.

\subsection{Resilience to errors: the 1D Heisenberg model}
\begin{figure}[t]
\includegraphics[width=0.8\columnwidth]{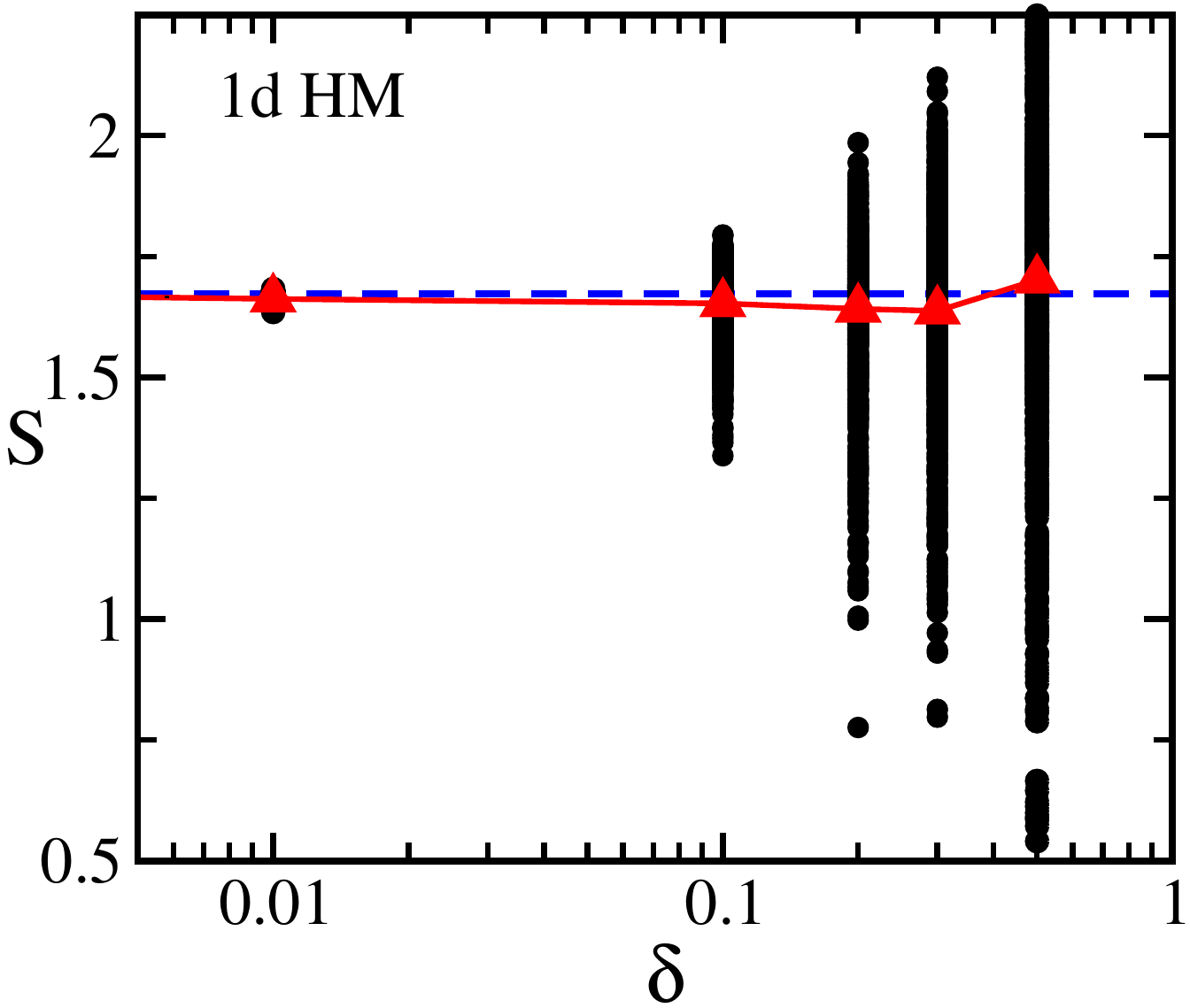}
\centering
\caption{Von Neumann entropy evaluated via the BW-EH as a function of the disorder magnitude $\delta$ for the
1D HM with L = 16 (see text). We consider the infinite PBC partition. 
The circles (black points) are the value of $S$ for a single realization of disorder, while
the triangles (red points) are the averaged $S$ ($N_r =$ [100 - 200] realizations of disorder are used).
The horizontal dashed line represents the value of $S$ in the clean case.}
\label{fig:1dstability}
\end{figure} 

We now consider the effect of disordered couplings, $\Gamma(n) \to \Gamma(n)(1+\delta_n)$, where $\delta_n = [-\delta,\delta]$, in the BW-EH of the 1D HM. Specifically, we are interested in understanding how the BW VNE is affected by a small amount of disorder, which is an important issue in an experimental context.

In Fig.~\ref{fig:1dstability} we show that, similarly to the 2D case [see Fig. 4 (b1) of the main text], the BW VNE is not appreciably affected by disorder up to strength of the order of  $10\%$. 
For larger values of $\delta$, we observe a considerable dependence on the disorder realization, as signalled by the visually large spreading of the values of $S$. Surprisingly, 
the mean value of the entropy is not dramatically affected.

\section{Bisognano-Wichmann entanglement Hamiltonian for 2D systems}

\begin{figure}[t]
\includegraphics[width=0.99\columnwidth]{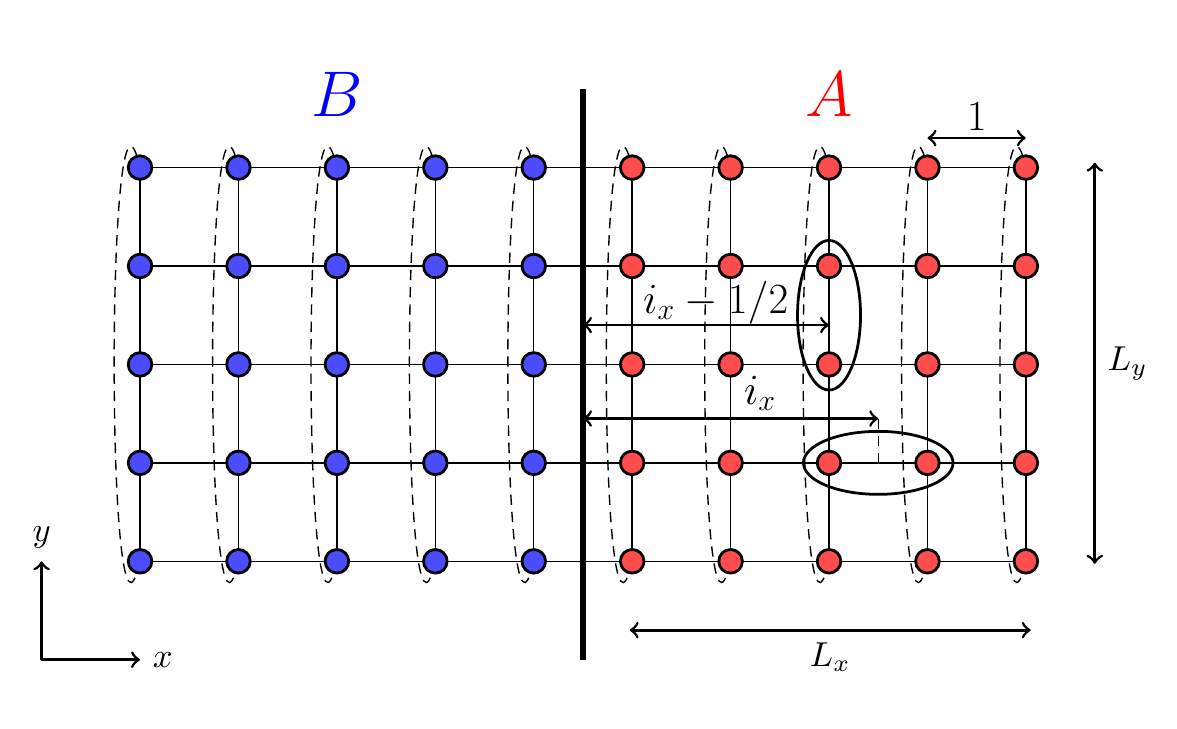}
\centering
\caption{Sketch of the two-dimensional system considered in this work.
The BW-EH is defined in the half-bipartition $A$.}
\label{fig:2d}
\end{figure}

We now review how to cast the  BW-EH on two dimensional lattices~\cite{Giuliano2018}.
As a concrete example, we consider the 2D Heisenberg model in a square lattice $L_x \times L_y$.
In this case, the BW-EH is
\begin{eqnarray} \label{xxz2D}
H_\textup{BW}&=&  \sum_{\vec{i},\delta = \pm 1} \Gamma \left( {i}_x \right) S_{(i_x,i_y)} S_{(i_x+\delta,i_y)}  \nonumber\\  
&+&  \sum_{\vec{i},\delta=\pm 1} \Gamma \left({i}_x - 1/2 \right) S_{(i_x,i_y)} S_{(i_x,i_y+\delta)},
\end{eqnarray}
where the lattice spacing has been set to 1 without loss of generality. The simulation of the subsystem BW-EH is performed considering periodic boundary condition in the $y$ direction, and 
open boundary condition in the $x$ direction, see Fig. \ref{fig:2d}.
The function $\Gamma(x)$ is given by the BW theorem
\begin{eqnarray}
\Gamma \left( x \right) = x,
  \label{BW}
\end{eqnarray}
which represents the EH of a half-bipartition;  we call this subsystem-geometry of cylinder.
Furthermore, we consider 
\begin{eqnarray}
 \Gamma \left( x \right) = \frac{x(L-x)}{L}, 
  \label{CFT}
\end{eqnarray}
which corresponds to the generalization of the BW to a subsytem that is embbeded in a infinite system; we call this subsystem-geometry of toroid.

We remind the reader that, as discussed in Ref.~\cite{Giuliano2018}, for finite values of $L_y$, formula Eq.~\eqref{CFT}
%[{\color{red}these formulations}] 
is in principle only applicable to conformal field theories. In order to mitigate this effect, we perform systematically extrapolations to 
large $L_y$ when determining universal properties. Let us illustrate here a simple, non-rigorous argument that partly justifies the applicability of this approach to generic (i.e., non conformal) 
2D models. Typically, the low energy theory will be made of gapless and gapped sectors. The description of the former will be scale invariant and relativistic invariant:
while this does not guarantee emergent conformal invariance, exceptions are rare. The gapped part of the theory will (at most) contribute to the entanglement properties only in the very vicinity of 
the edge of the partition, where it would actually behave like a gapless theory. Far from the boundary, the reduced density matrix with respect to these degrees of freedom will be an identity operator (up to degeneracies). 
This indicates that the CFT formulas used above shall be applicable also to more general cases where some low-energy degrees of freedom are actually gapped. In the context of the 2D HM, the role of gapless degrees
of freedom is played by the $\mathbb{C}P(1)$ model describing the emergent Nambu-Goldstone modes, and the gapped part of the theory is described by the massive Goldstone mode.

\section{Quantum Wang-Landau sampling of the  entanglement Hamiltonian}

In this section we discuss some relevant details of the quantum Wang-Landau (WL) simulations used in this work.
Compared with the convetional quantum Monte Carlo (QMC) simulations, that is performed at a fixed temperature, the WL method features two main advantages
for the study of the thermodynamic properties of the EH:
(i) it allows to directly compute the thermal entropy at the ``entanglement temperature'' $\beta_{EH}$,
and (ii) the thermodynamic properties of the EH are obtained for a broad range of temperature with a single run of the simulation.

The WL method was originally proposed for classical systems in Ref.~\cite{Wang2001}.
The key idea of the method is to calculate the density of states, $\rho(E)$.
For a quantum Hamiltonian, such as the BW-EH, however, one must map the system to a classical one.
This is done, for instance, using the  SSE framework,
which considers the following form for the partition function
\begin{equation}
 Z = \tr{e^{-\beta H}} = \sum_{n=0}^{\infty} \frac{\beta^n}{n!} \tr{\left( - H \right)^n} =  \sum_{n=0}^{\infty} \beta^{n} g(n),
 \label{Z}
\end{equation}
where the \textit{n}th order series coefficient $g(n)$ plays the role of the density of states in the classical algorithm.
We refer  to Ref.~\cite{Wessel2003,Wessel2007} for the general details of the computation of $g(n)$.
Below we mention the technical aspects of the simulation that are relevant to reproduce our results.

%An important aspect of the simulation is that 
The series expansion Eq.\eqref{Z} can be
truncated at an order $\Lambda$, i.e., $n = 0,1, ..., \Lambda $, without introducing systematic errors in the simulation.
The choice of $\Lambda$ is performed as in the conventional SSE simulations, see Refs. \cite{sandvik1991,sandvik2002},
which  gives as a result  $\Lambda(\beta) \approx \beta |E(\beta)|$; where $E(\beta)$ is the expectation value of the total energy at inverse temperature $\beta$.
The effect of the cutoff $\Lambda(\beta)$ is that the range of temperature that can be accessed is restricted to $\beta < \Lambda(\beta)/E(\beta)$.
In order to obtain the results of Fig. 4(a,b1) of the main text, for instance, we simulate the BW-EH using  $\Lambda(3\beta_{EH})$ as the cutoff.
Instead, the computation of the BW VNE at $\beta_{EH}$ are obtained utilizing a cutoff $\Lambda(\alpha \beta_{EH})$.
The results shown in Figs.~2  and 3 of the main text are obtained with $\alpha = 1.3$. 
We  check that these results don't change upon increasing $\alpha$.

In the WL-SSE algorithm, the sampling of the SSE configurations with different  \textit{n} is performed  with
a probability function that is proportional to the inverse of the ``density of states``, $1/g(n)$.
The WL sampling generates a histogram for the distribution of \textit{n} that is flat, i.e., $H(n) \sim const$;
the histogram $H(n)$ is obtained counting the number of times a configuration with \textit{n} is observed.
The key point of the algorithm is that $g(n)$ can be computed  by iteratively flattening $H(n)$.
More specifically, one start with the guess  $g(n) = 1$.
Further, each time the configuration  \textit{n} is accepted $g(n)$ is multiplied by a factor $f$, i.e., $g(n) \to g_{old}(n)f$.
This proccess is repeated until $H(n)$ is flat.
In practice, we consider as a condition for the flatness of $H(n)$ a maximum deviation of 20$\%$  from the mean value.
Once $H(n)$ is flat, it is reset to zero, and $f$ is decreased by  $\ln(f) \to \ln(f_{old})/2$  \cite{Melko2013}. 
This proccess is repeated until convergence is achieved.
Here we use the  convergence condition proposed in  Refs. \cite{Melko2013,Pereyra2008}.

In addition to the aforementioned algorithm, we consider the optimized-broad-histogram algorithm proposed in Ref. \cite{Wessel2007}
for the 2D Heisenberg model, see Fig.3 (a) of the main text.
These results were obtained with the  ALPS code \cite{Bauer2011,codeALPS}
In this case, we confirm that the two methods give the same results (within error bars).

Finally, it is important to mention that the results of the BW entropy are obained by doing an average of $N_r$ independent
WL simulations, i.e.,
\begin{equation}
 S(\beta) = \frac{1}{N_r} \sum_{i=1}^{N_r} S_i(\beta).
\end{equation}
The error bars are the standard deviation of the distribution $\{S_i\}$, and for all the results presented, we consider at least $N_r > 200$.

\end{document}